\documentclass[pre,12pt,
               superscriptaddress,
               onecolumn,
               showpacs,
               ]{revtex4-1}
               
\usepackage{graphicx}   
\usepackage{dcolumn}    
\usepackage{bm}         
\usepackage{amsmath}
\usepackage{amssymb}
\usepackage[usenames]{color}      

  \usepackage[breaklinks=true, colorlinks=true, citecolor=blue, linkcolor=blue, urlcolor=blue, bookmarksdepth=3, pdftitle={Bubble puzzles: From fundamentals to applications}]{hyperref}
  
\usepackage{fancyheadings}
\def\be{\begin{equation}}
\def\ee{\end{equation}}
\newcommand{\ba}{\begin{eqnarray}}
\newcommand{\ea}{\end{eqnarray}}

\def\x{{\mbox{\boldmath$x$}}}
\def\y{{\mbox{\boldmath$y$}}}
\def\u{{\mbox{\boldmath$u$}}}

\def\f{{\mbox{\boldmath$f$}}}

\def\v{{\mbox{\boldmath$v$}}}
\def\g{{\mbox{\boldmath$g$}}}
\def\bomega{{\mbox{\boldmath$\omega$}}}
\def\calv{{\cal V}_b}

\begin{document}

\title{Bubble puzzles: From fundamentals to applications}

\author{Detlef Lohse}
\affiliation{Physics of Fluids Group, Max-Planck Center for Complex Fluid Dynamics,
MESA+ Research Institute and J.M. Burgers Centre for Fluid Dynamics, University of Twente, P.O. Box 217, 7500 AE Enschede, The Netherlands.}
\affiliation{Max Planck Institute for Dynamics and Self-Organization, Am Fassberg 17,
37077 G\"ottingen, Germany}

\date{\today}

\begin{abstract}
For centuries, bubbles have fascinated artists, 
engineers, and scientists  alike. In spite
of century-long research on them, new and often surprising 
bubble phenomena, features, and applications
keep popping up. In this paper I sketch my personal scientific bubble journey, starting
with single bubble sonoluminescence, continuing with sound emission and scattering of bubbles,
 cavitation, snapping shrimp, impact events, 
air entrainment, surface micro- and nanobubbles,  
 and finally coming to effective force models for bubbles and 
 dispersed bubbly two-phase flow. In particular,  I also cover 
various applications of bubbles, namely in ultrasound diagnostics,
drug and gene delivery, 
piezo-acoustic inkjet printing, immersion lithography, sonochemistry,
electrolysis, catalysis, acoustic marine geophysical survey, and bubble drag reduction for naval vessels,  
and show how these applications crossed my way. I also try to 
show that good and interesting fundamental 
science and relevant applications are not a contradiction, but mutually stimulate each other
in both directions. 
\\ 
\\ 
\noindent DOI: \href{https://doi.org/10.1103/PhysRevFluids.3.110504}{10.1103/PhysRevFluids.3.110504} 
\end{abstract}

\maketitle
\thispagestyle{fancy} 

\section{Introduction}\label{intro}
``How do you find the problems you work on?" And ``What do you think is the 
difference between fundamental  and applied research?''
These are questions I am often asked. My short answer to the first  question is: ``Be curious!''
And to the second one: 
``In principle, none''. 
And what in particular holds for both fundamental and applied problems, both in finding and in solving them: ``Watch,
 listen,  and be open''. 
 The answer to both questions can be summarized as: 
``Work on problems you most enjoy. Strange things can happen on the  way.'' (Walter Munk, UCSD). 
In the best case, the problem to work on is both relevant and outstanding, at the same time.

In this article 
I want to take the opportunity to give longer answers to these questions, in particular by giving 
examples from my own scientific biography and scientific journey. As the thread of the article I will choose ``bubbles''. 
I will report how I first incidentally bumped into the science of bubbles, what and how I learned  about 
 them, what 
wonderful science and great interactions and collaborations
 with colleagues this endeavor opened for me, and how I kept on bumping into very relevant applications of bubbles
in technology. So bubbles have provided me both wonderful scientific problems and very relevant applied questions,
to whose solution, I think, we have  contributed over the last two and a half decades. 

The length scales on which bubbles are relevant  range from nanometers to at least tens of meters, and I will give 
examples for interesting and relevant bubble phenomena on 
all these scales. 
The richness of bubble fluid dynamics is reflected in the many dimensionless numbers that  are relevant
in the context of bubbles \cite{bre95,lei94}, namely the
\begin{itemize}
\item
Reynolds number $Re = U R/\nu$, expressing the ratio of inertia forces to viscous forces. 
Here $U$ is the bubble velocity, $R$ its radius, and $\nu$ the kinematic viscosity.
\item
Froude number $Fr = U/ \sqrt{g R}$,  ratio of inertia to buoyancy, where $g$ is the gravitational
acceleration. 
\item
Archimedes number $Ar = g R^3 \rho (\rho-\rho_g) /\eta^2$ , ratio of buoyancy to viscosity.
$\rho$ is the liquid density, $\eta$ its  dynamic viscosity, and $\rho_g$ the gas density. 
\item
Galileo number $Ga = g R^3/\nu^2$, ratio of gravitational to viscous forces,
\item
Weber number $We = \rho U^2 R/\sigma$, ratio of inertia to capillarity, where $\sigma$ is the surface tension. 
\item
Capillary number $Ca = \eta U/\sigma$, ratio of viscous  to capillary forces,
\item Ohnesorge number $Oh= \eta /\sqrt{\rho \sigma R} = We^{1/2} / Re$, ratio of time of viscous damping  to time  of the capillary oscillations. 
\item
E\"otv\"os number $Eo  = (\rho- \rho_g) g R^2 /\sigma$,  also called Bond number $Bo$,
 ratio of buoyancy to capillarity,
\item Stokes number St, ratio of characteristic timescale  of bubble to that of the flow, 
\item Morton number $Mo = g \eta^4 (\rho - \rho_g) / (\rho^2 \sigma^3) $, 
which is a material parameter for a bubble in a certain liquid,
depending only on surface tension, density, density contrast, viscosity, and gravity.
\item Lewis number $Le = \kappa / D$, ratio of thermal diffusivity $\kappa$ 
to mass diffusivity $D$, and thus another material parameter, 
\item Damk\"ohler number Da, ratio between  chemical reaction rate and 
(diffusive or convective) mass transport
rate,
\item Jakob number $Ja = \rho c_p (T - T_{sat}) / (\rho_v \Lambda) $, ratio of sensible heat to latent heat. Here $c_p$ is the liquid specific heat, $\Lambda $ is the latent heat, $\rho_v$ the vapor density,
$T_{sat}$ the saturation temperature of the liquid, and $T$ the temperature of the  surrounding
liquid. 
\item ....
\end{itemize}
We will encounter most of these numbers in this article.

The selection of which bubble problems I will report on is naturally subjective and, as said above, along my own 
scientific 
bubble journey. 
This will also be reflected in the citations, where I will restrict myself to papers that  had a 
significant  scientific impact on me and to  references to our original work. 
Some of the given examples I briefly 
discussed before, in a  short Proceedings \cite{loh06p} (without
Web of Science Index). 
There are many more bubble problems and applications, which I cannot report here or which I am even not aware of,
but I hope that the paper stimulates other scientists to look into the subject and be open towards both fundamental 
and applied bubble problems, because -- as I hope to be able to show with this article --
 it is intellectually very rewarding, covers many areas of fluid dynamics,  and extremely relevant in many applications.

\section{Sonoluminescence -- illuminated bubble dynamics}\label{sl}
\begin{figure}[h]
\begin{center}
\begin{minipage}{0.5\linewidth}
\includegraphics[width=0.98\textwidth]{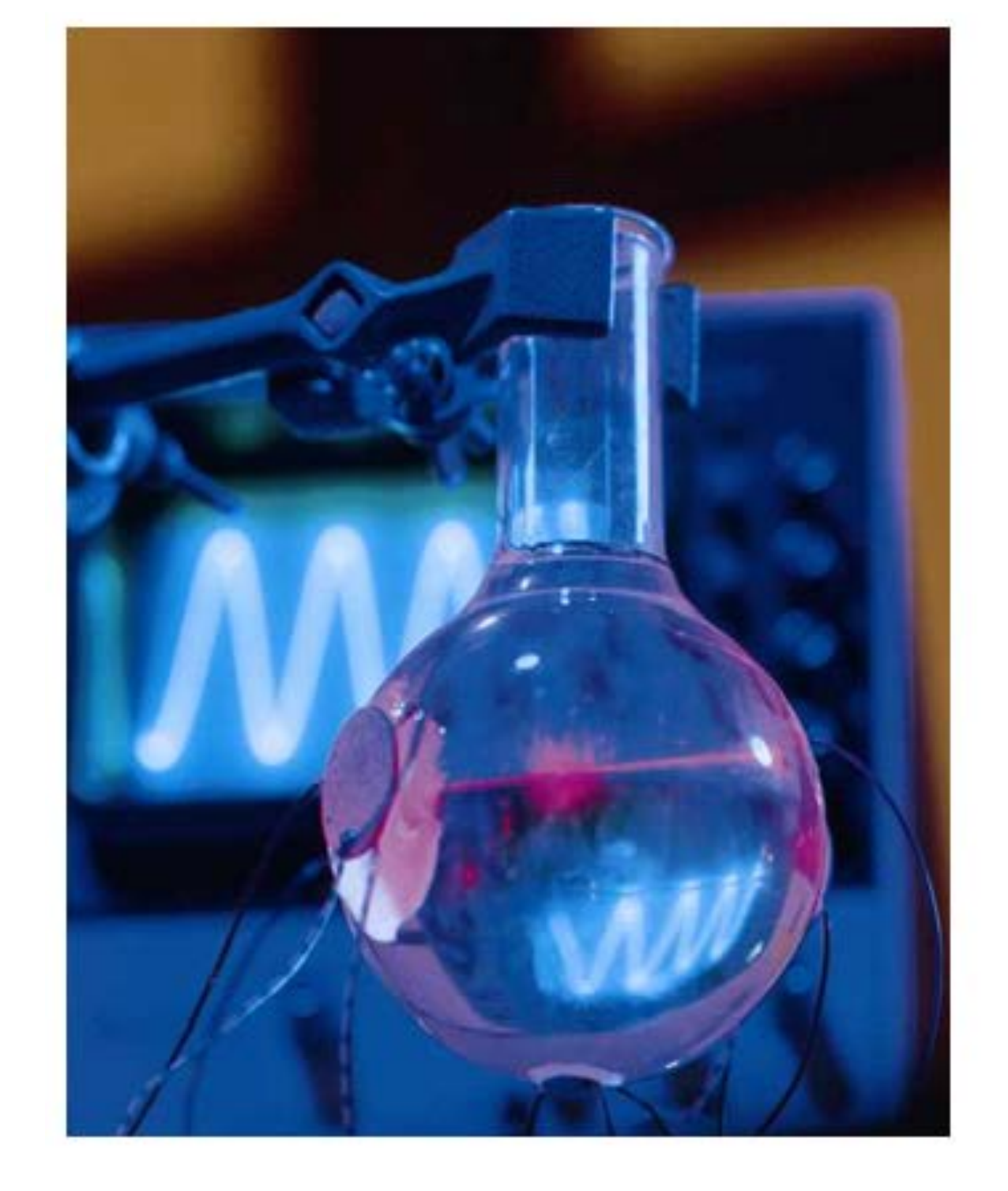} 
\end{minipage}\hfill
\begin{minipage}{0.45\linewidth}
\caption{{  \it 
Setup for single bubble sonoluminescence: Piezoelectric transducers  are
glued to a flask filled with water. They excite a standing acoustic
wave in which the light-emitting bubble is  trapped. Photo taken by R\"udiger Toegel,
Physics of Fluids, Twente, 2000. 
    }}
\label{toegel}
\end{minipage}
\end{center}
\end{figure}

\begin{figure}[htb]
   \centering
\includegraphics[width=0.96\textwidth,angle=-0]{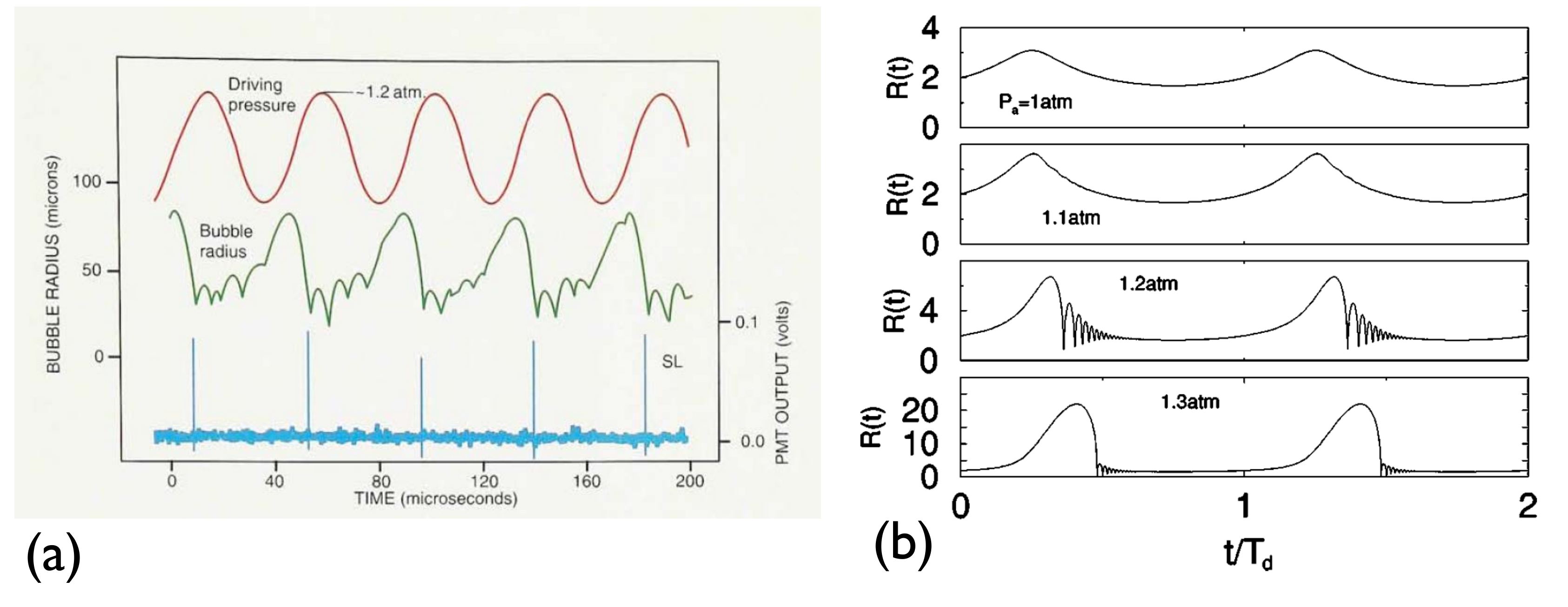}
\caption{\it
(a) Acoustic driving pressure $P(t)$ (in red), 
resulting bubble  radius $R(t)$ (in green), 
and light intensity $I(t)$ (in blue), as measured by
\cite{gai92}.  A negative driving pressure causes the bubble
to expand; when the driving pressure changes sign, the bubble collapses,
resulting in a short pulse of light (marked SL).
The figure is taken from ref.\ \cite{cru94b}.
(b)
Solutions to the Rayleigh--Plesset equation (\ref{rp}) with a sinusoidal driving
$P(t) = P_a \sin (\omega t)$ 
at forcing pressures $P_a=1.0,1.1,1.2,1.3$\,atm.
The ambient bubble radius is $R_0=2\mu m$ and the frequency $f= \omega/(2\pi) =26.5$ kHz.
Material parameters are for water at room temperature. 
}
\label{fig-sl-dyn}
\end{figure}

The first major scientific bubble problem I bumped into was sonoluminescence. In 1994, during my time
as postdoc at the University of Chicago in the group of Leo Kadanoff, I attended a lecture
by Brad Barber on his PhD thesis on single bubble sonoluminescence \cite{barber1992,bar91b} (later summarized in 
\cite{bar97}). This phenomenon had been discovered a few years earlier by Felipe Gaitan \cite{gai90,cru94b},
then a PhD student in Mississippi, when he experimented with
an air bubble trapped in a water-filled flask by piezo-acoustical forces (the so-call Bjerknes force \cite{bje09,pro77}, see figure \ref{toegel} for a similar setup), 
which at the same time drive the bubble:
When the pressure is low, the bubble expands, and once it is high, it is compressed. But what 
Gaitan observed was first not believed by anybody:  
Under certain conditions the bubble can emit  light! How can this be? Typical
acoustical energies are in the range of $10^{-12}$ eV per molecule,
typical light energies in the range of 1 eV. This means that there is an
energy focusing factor of $10^{12}$! Sound, radius, and light intensity as function of time
are reproduced in figure \ref{fig-sl-dyn}a.


Directly after Barber's talk, I discussed this fascinating subject with my
colleague Michael Brenner, then also a postdoc at the University of Chicago. 
We asked ourselves the two obvious questions: 
 What is the light emitting process and
under what conditions does this phenomenon happen, i.e., what is the
phase space of single bubble sonoluminescence? We first focused on the second question, and started to 
read and learn about bubble dynamics.

We very soon found the seminal papers by Andrea Prosperetti on this subject, 
most visibly summarized in his review \cite{ple77b}. The core dynamical 
equation is the celebrated Rayleigh-Plesset (RP)
equation for the bubble radius $R(t)$,
\begin{equation}
R \ddot R + \frac{3}{2} \dot R^2 = \frac{1}{\rho}\left( p_{g} - P_0 - P(t)
- 4 \eta \frac{\dot R}{R}- \frac{2 \sigma}{R} \right),
\label{rp}
\end{equation}
where $p_g(R(t))$ is the pressure inside the gas bubble, 
$P(t) = P_a \sin (\omega t)$ the driving acoustical pressure with amplitude $P_a$ and frequency 
$f= \omega/(2\pi)$, $P_0$ the ambient pressure, $\eta$ the dynamic viscosity, and $\sigma $ the surface tension.
  A historical review of the development of this equation is
given in ref.\  \cite{ple77b}. The typical RP bubble dynamics for increasing driving pressure
$P_a$ is shown in figure \ref{fig-sl-dyn}b.

The  left-hand side of the RP equation (\ref{rp})
 was already  known to  Lord Rayleigh who derived  it in the context of an  analysis of
 cavitation damage of  ship propellors \cite{ray17,bre95,loh03a}. So already in Rayleigh's time  applied and fundamental
 science came hand in hand. The solution to the  inertial part $R \ddot R + \frac{3}{2} \dot R^2 = 0$
 of the RP equation (\ref{rp}) is the power law
 \begin{equation}
 R(t) \propto (t_s - t)^{2/5},
 \label{sing}
 \end{equation}
 with a diverging singularity in the bubble wall velocity $\dot R(t) \propto (t_s - t)^{-3/5}$ at time $t_s$.
 This singularity reflects the violent bubble collapse which can occur for strong enough driving,
 see figure \ref{fig-sl-dyn}. In a nutshell, at collapse, the gas inside the bubble gets compressed,
 heats up, partly ionizes, and at recombination light is emitted \cite{cru94b,bre02}.

The work of Andrea Prosperetti also led us to the conditions under which stable single bubble 
sonoluminescence can occur: An obvious necessary condition is the (spherical) shape stability 
of the bubble, which Eller and Crum \cite{ell70} 
had experimentally  and Prosperetti  theoretically analysed 
 \cite{pro77b}. We applied his results and determined  under what conditions the collapsing
 bubble would be (spherical) shape stable so that it would on the one hand survive the collapse, but at the 
 same time would still collapse strongly enough so that the gas inside the bubble would be 
 considerably heated \cite{bre95b}. 
 Both a parametric instability and the Rayleigh-Taylor instability turned out to be relevant
 \cite{bre95b,hil96,bre02}.
 
 Another necessary condition for stable single bubble
 sonoluminescence  is the diffusive stability
 of the bubble. Also  the diffusive bubble stability 
 had been analysed before, namely in the seminal work by 
 Epstein and Plesset \cite{epstein1950},  later extended by Fyrillas and Szeri \cite{fyr94} to oscillating 
 bubbles, for which rectified diffusion \cite{bla49b,ell64,cru80}
 can occur: For large pressure, the bubble loses gas to the 
 outside liquid,
 but during the low pressure period, it can gain gas from outside. For very strong driving 
 the growth can win, mainly due to the thin boundary layer during that time but also due to the 
 much larger bubble size. 
 Applying these ideas  to the regime of sonoluminescing bubbles  could account for the experimentally
 observed diffusively 
 stable single bubble sonoluminescence \cite{bre96b}, and combining the conditions of shape 
 stability, diffusive stability, and energy focusing 
  led to the phase diagram of sonoluminescing bubbles
 \cite{hil96}, which for pure argon bubbles was in good agreement with the experimental
 observations. 
 
\begin{figure}[htb]
   \centering
\includegraphics[width=0.99\textwidth]{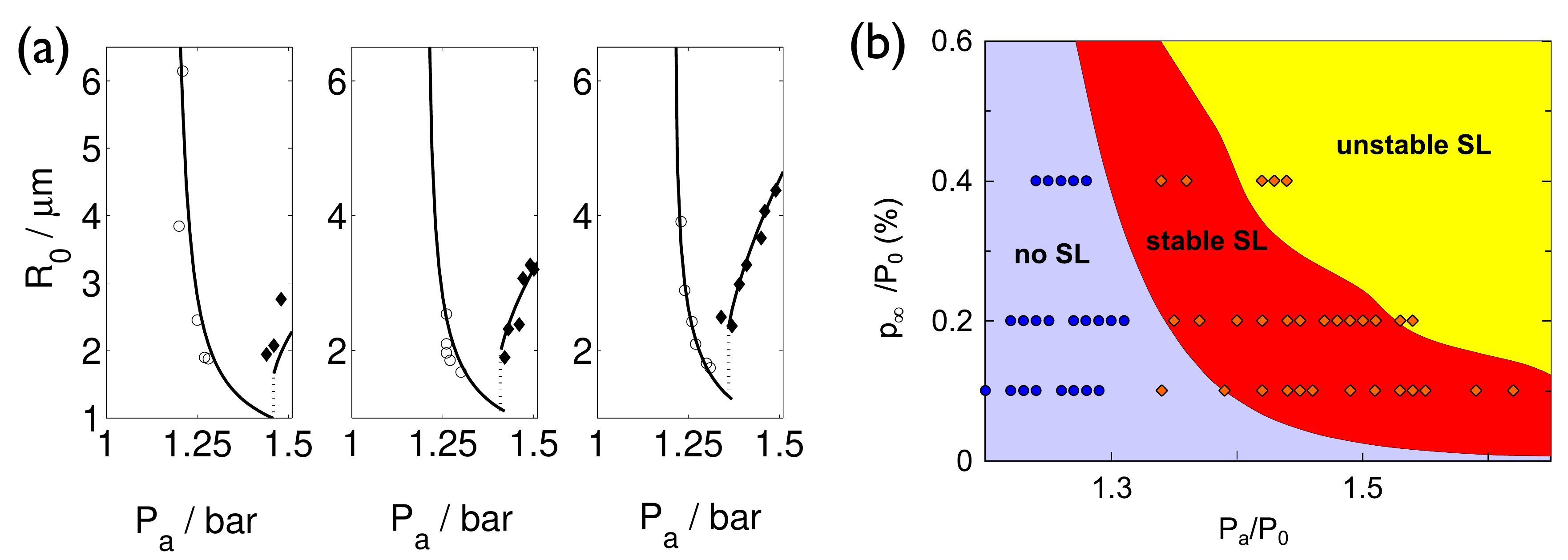}
\caption{\it
(a) Phase space of single bubble sonoluminescence in the argon  concentration $P_\infty / P_0$ vs.\
driving pressure $P_a/P_0$ phase plane for pure argon bubbles, showing the three phases
``no SL'', ``stable SL'' (only possible for very small argon concentrations), and ``unstable SL'', where the bubbles
grow by rectified diffusion while emitting light and finally run into  a shape instability \cite{hil96}. 
Adopted from \cite{hil96}. 
(b) Phase
space  of single bubble sonoluminescence in the bubble radius vs.\ driving pressure 
phase plane
for air bubbles, taken from ref.\
\cite{toe03}:
Bubble equilibrium radius $R_0$ versus driving pressure $P_a$ for
three different air concentrations (10\%, 20\%, and 40\% of saturation;
the driving frequency is 33.4 kHz). The curves, signalling stable bubbles,  follow from our parameter-free
theory \cite{toe03};
the data points had been measured by Ketterling and Apfel \cite{ket98}.
The bubble emits light
only on the right curves where argon has accumulated in the bubble (filled diamonds).
On the left curves,  the losses by chemical reactions and the growth by rectified diffusion balance. 
}
\label{fig-sl-photo}
\end{figure}

 However, air bubbles were found to be stable for 100 times larger gas saturation than pure argon
 bubbles, see figure \ref{fig-sl-photo}. The reason for this turned out to be 
 the {\it chemical stability}  of the gas inside the bubble \cite{loh97}:
The bubble is collapsing  so strongly
that the gas inside is nearly adiabatically compressed. This means that the
collapse of the bubble is so violent that no thermal equilibrium with
the surrounding water can be established: The bubble is heating up,
to about 15000 K, as we now know both theoretically \cite{toe03}
and experimentally \cite{fla05}. For molecular gases such as O$_2$ or
N$_2$ this is  much too hot and they dissociate: The resulting
radicals react with each other and with the dissociation products of
water vapor.  NO, NH, etc.\
 are formed,  which dissolve in water. Therefore it is mainly
argon that  remains in the bubble, which is contained in air with a concentration
of about 1\%, explaining the factor of 100 higher gas concentration required for stable single
bubble sonoluminescene with 
air as compared to the pure argon (or any other inert gas) case.  
This theory later got confirmed through  various experimental results, see e.g.\  \cite{mat98,ket98}. 
Another advantage of argon for achieving strong SBSL  is that  
in contrast to O$_2$ or N$_2$,
argon has no internal degrees of freedom. Thus the focused  energy of
the bubble collapse can directly be transferred into heat. The
15000 K which are reached in this way are sufficient to partly ionize
the gas. Recombination of ions and electrons leads to light emission through thermal 
bremsstrahlung
\cite{mos97b,hil99b}. 
Later we extended the ideas of chemical stability of ref.\ \cite{loh97} to include various other
chemical reactions
of air with water \cite{toe03} to find very good and quantitative agreement between
experimental and theoretical phase diagrams, see figure \ref{fig-sl-photo}b.

In a sense,  single bubble sonoluminescence can be viewed
 as {\it illuminated bubble dynamics}, with the RP dynamics as backbone. 
A combination of concepts from
hydrodynamics (both shape stability and diffusive stability), 
chemistry, plasma physics, applied mathematics,
thermodynamics, and acoustics led to the phase diagrams,  which are in good agreement with
the experiments \cite{bre02,toe03}. In its conceptual simplicity -- 
an isolated, fixed, non-interacting  single bubble in a flask -- 
it can also be seen as ``hydrogen atom of bubble fluid dynamics'', on which we learned  a lot.

Our work on sonoluminescence started off as pure fundamental research,
driven by curiosity. We had not asked ourselves  whether there would be
 any applications. Single bubble sonoluminescence simply was a fascinating
  and outstanding problem,
 with major open questions. In answering them, we learned tremendously, including on 
 \begin{enumerate}
 \item Acoustic and other forces on bubbles 
 \item Bubble dynamics  and bubble shape stability 
 \item Diffusive bubble stability 
 \item Chemical bubble stability 
 \item Bubble nucleation 
 \item Bubble collapse and cavitation 
 \item Plasma formation and thermal bremsstrahlung 
 \end{enumerate}
As we will see, all these items turned out to be 
very relevant in connection with various applications -- sometimes  very unexpectedly. 
The insight obtained from single bubble sonoluminescence therefore helped us enormously 
to discover and identify applied problems and to help to solve them or at least make some progress
on them. In the next chapters I will report how this came about. 

Namely, I will report on:
\begin{itemize}
\item How  our understanding of the bubble dynamics and of the bubble shape stability 
(subject  2 in above list) 
contributed to ultrasound diagnostics and to improve ultrasound contrast agents. I will also report
on other applications of bubble dynamics in the medical context (section \ref{med}). 
\item 
How we found out that the sound of snapping shrimp originates from a cavitating bubble 
(section \ref{shrimp}, originating from subjects  5 and 6 in the above list).
\item How the collapse of a bubble or a void close to a  surface focuses the energy, leading
to a major jet (section \ref{impact}, originating from subject  6 in the above list).
\item 
How an entrained bubble in a  piezo-acoustic inkjet channel can cause major trouble
due to rectified diffusion (subject 3 in the above list) and how to solve this problem (section \ref{inkjet}).
\item How bubbles can nucleate on a microstructured  
surface and, when acoustically driven, collapse in 
a controlled way,  enhancing  the efficiency of ultrasonic cleaning and chemical reactions 
(section \ref{sonochemistry}, originating from subjects  4, 5 and 6 in the above list).
\item How our understanding of diffusive bubble stability (subject 3 in the above list)  was instrumental
to figure out why surface nanobubbles and surface nanodroplets are stable, with various applications
in electrolysis, catalysis, diagnostics,
 and the food and remediation industry (section \ref{nb}). 
\item Finally, how our understanding on bubble forces (subject 1 in above list)  brought us to 
bubbly two-phase flow, including studying drag reduction in turbulent bubbly flow, for which
bubble deformability (subject 2) is crucial (section \ref{force}). 
\end{itemize}
The paper closes with conclusions and with a short 
outlook (section \ref{conclu}). In particular, I will motivate
why from my point of view we live in the golden age of fluid dynamics. Both wonderful bubble science and 
very relevant bubble applications are ahead of us.

\section{Ultrasound diagnostics, ultrasound contrast agents, and other applications of bubbles in diagnostics, therapy, and medicine}\label{med}

\begin{figure}[htb]
\centering
\includegraphics[width=0.98\textwidth]{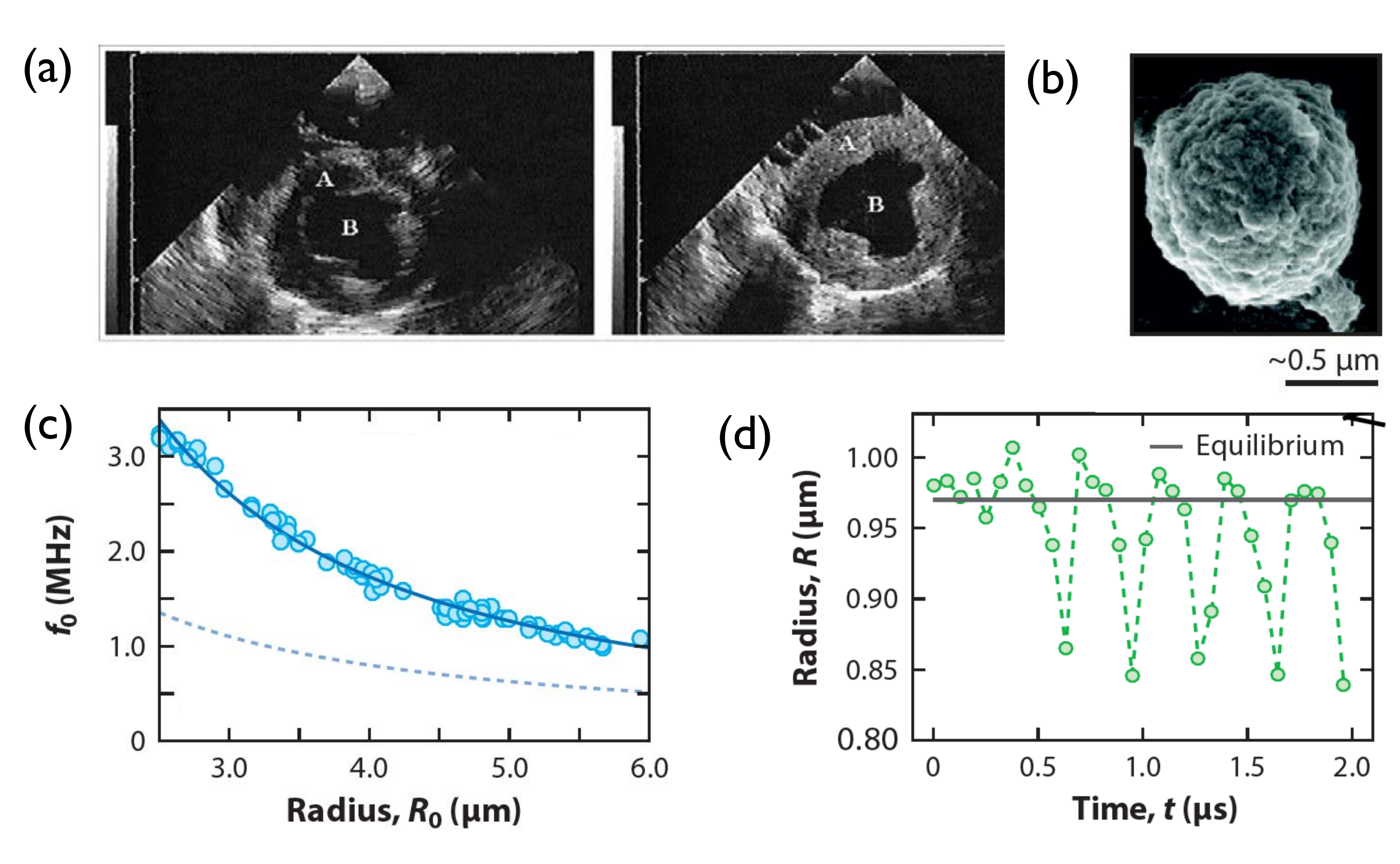}
\caption{
\it
(a) Ultrasound image of a heart without (left) and with (right) previous injection
of micro-bubbles. In the second case the structures become clearer as the
bubbles act as ultrasound contrast enhancers.
(b) Electron micrograph of a microbubble coated with a protein layer, taken from ref.\ \cite{cavalieri2008}. 
(c) Eigen frequency $f_0$ of a sound-driven coated with DPPC monolayers
 as function of the ambient bubble radius (blue data points, from \cite{lum2016}).
The solid blue line shows the fit to the Marmottant model \cite{mar05}, whereas the dashed line shows the Minnaert frequency
(\ref{rp-lin}). Figure adopted from ref.\ \cite{dollet2019}. 
(d) With the coating the sound-driven bubbles show the typical ``compression only'' behavior, with the compression being 
much more pronounced than the expansion. Figure taken from ref.\ \cite{mar05}.
}
\label{fig-uca}
\end{figure}

In an outreach effort, in order to popularize physics, fluid dynamics, and bubbles,
in 1995 I had
written an article on single bubble sonoluminescence in ``Physikalische Bl\"atter'', which was
the German analog of  Physics Today. Based on this, I got contacted by 
a physicist working for a pharmaceutical company on ultrasound contrast agents (UCA),
which are introduced into the blood to enhance the acoustic scattering 
and which 
contain  small, encapsulated microbubbles.  These 
very effectively  scatter ultrasound, see fig.\ \ref{fig-uca}a.
 In this way, it is e.g.\ possible to visualize the perfusion of tissue, like  the heart muscle. 
 The images  are meanwhile  
used to  obtain 
 diagnostic information from the volume, shape and movement of the heart ventricles, in studying the blood flow in small blood vessels,  in blood perfusion measurements, 
 and in targeted molecular imaging, 
 among others \cite{fri00,bec00,wilson2010}.

One of the nagging questions in ultrasound diagnostics in the mid 1990s was: How does one 
increase the signal-to-noise ratio? Namely, when detecting the emitted sound from the bubble at the driving frequency, the signal is obscured by the reflections from tissue. To improve the signal quality, it had  been proposed \cite{burns1996a} 
 to detect higher harmonics of the driving frequency in the sound emission spectrum of the bubble.
 Vice-versa, also subharmonics  had been suggested for better contrast. 
The immediate question was: What bubble properties are optimal for these purposes, in particular 
what bubble size, for given driving frequency? This indeed was the question with which the pharmaceutical 
industry approached us, and thanks to our work on single bubble sonoluminescence, we could straightforwardly
provide an
 answer, namely by simply solving the RP equation with the relevant parameter for medical 
ultrasound imaging: Here, rather than in the range of 20 kHz -- 30 kHz as common
 for single bubble sonoluminescence,
the 
  typical driving frequencies are  between 2 MHz and 10 MHz and typical 
bubble radii are a few micrometers. From linearizing the RP equation (\ref{rp}) it follows 
that the eigenfrequency of the volume oscillations of an acoustically driven  bubble 
with 
ambient radius $R_0$ and under isothermal conditions 
 approximately is \cite{ple77b,pro84b,bre95}
\begin{equation}
\omega_0  = \sqrt{  3 P_0 \over \rho_l R_0^2
}. 
\label{rp-lin}
\end{equation}
This eigenfrequency is called Minnaert frequency. 
With the material parameters for water under ambient conditions this gives the well-known rule-of-thumb
\cite{pro84b}
\begin{equation}
f_0  R_0 \approx 3 ~ \hbox{MHz} ~\mu\hbox{m}  = 3 ~ \hbox{kHz} ~\hbox{mm}  = 3 ~ \hbox{Hz} ~ 
\hbox{m}
\label{rp-res}
\end{equation}
for the resonance frequency $f_0 = \omega_0 / (2\pi)$. 
For the frequencies of medical ultrasound imaging, the resonance radii are thus in the micron range.
Also in the fully nonlinear case, thanks to the full RP equation 
we could calculate optimal parameter values for maximal sound emission in the 
second  harmonic and in  subharmonics, and could make statements on the expected 
bubble shape stability in those regimes
\cite{gro97_zom}. 

One issue we had  first ignored was that the 
ultrasound contrast agent bubbles are not ``naked'', but coated 
 with lipids and
polymers (figure \ref{fig-uca}b), to avoid the obviously undesirable bubble clustering in the body and to increase their lifetime. 
The coating however modifies the oscillation behavior of
the bubble in an {\it a priori} unknown way. 
Therefore, a few years later, to take the effect of the coating on the bubble dynamics into consideration, 
we developed  a model  to
quantitatively describe this modification of the RP dynamics (\ref{rp}), which is now known as ``Marmottant model'' \cite{mar05}.
The key idea behind this model is to introduce an instantaneous
 bubble-size dependent surface tension, reflecting the buckling
behavior of the bubble coating when compressing the bubble, an elastic regime, and a shell-ruptured regime. 
An excellent recent review of such modified  bubble dynamics 
can be found in ref.\ \cite{dollet2019}.

To experimentally test such models for the relevant frequencies in the MHz regime, one unavoidably needs 
ultra high-speed imaging, with frame rates $\gg$ 1 MHz. 
In an effort led by Nico de Jong and Michel Versluis, 
we therefore  developed \cite{chi03,gelderblom2012}  an ultrafast
camera, which allows  imaging 128 digital frames with a frame rate of
up to 25 MHz. We called it ``Brandaris 128'', as it is based on a
rotating mirror, just as the famous Dutch lighthouse Brandaris on
Terschelling. This camera allowed us to gain insight into the volume
and shape oscillations of ultrasound contrast agent bubbles \cite{mee07,dol08,overvelde2010}, in particular when combining it  
 with measurements of the
acoustic emission of such coated bubbles \cite{mee07}. This procedure allowed us to 
adjust the model parameters of the Marmottant model \cite{mar05} to the experimental data, very nicely reflecting 
the observed so-called ``compression-only'' behavior of the UCA bubbles \cite{mar05,mee07,jong2007,sijl2011} (see figures \ref{fig-uca}c,d), which
is nothing else than -- thanks to the bubble coating -- modified RP dynamics, thus giving the ultrasound contrast agent 
community a very relevant tool. 

\begin{figure}[htb]
\centering
\includegraphics[width=0.98\textwidth]{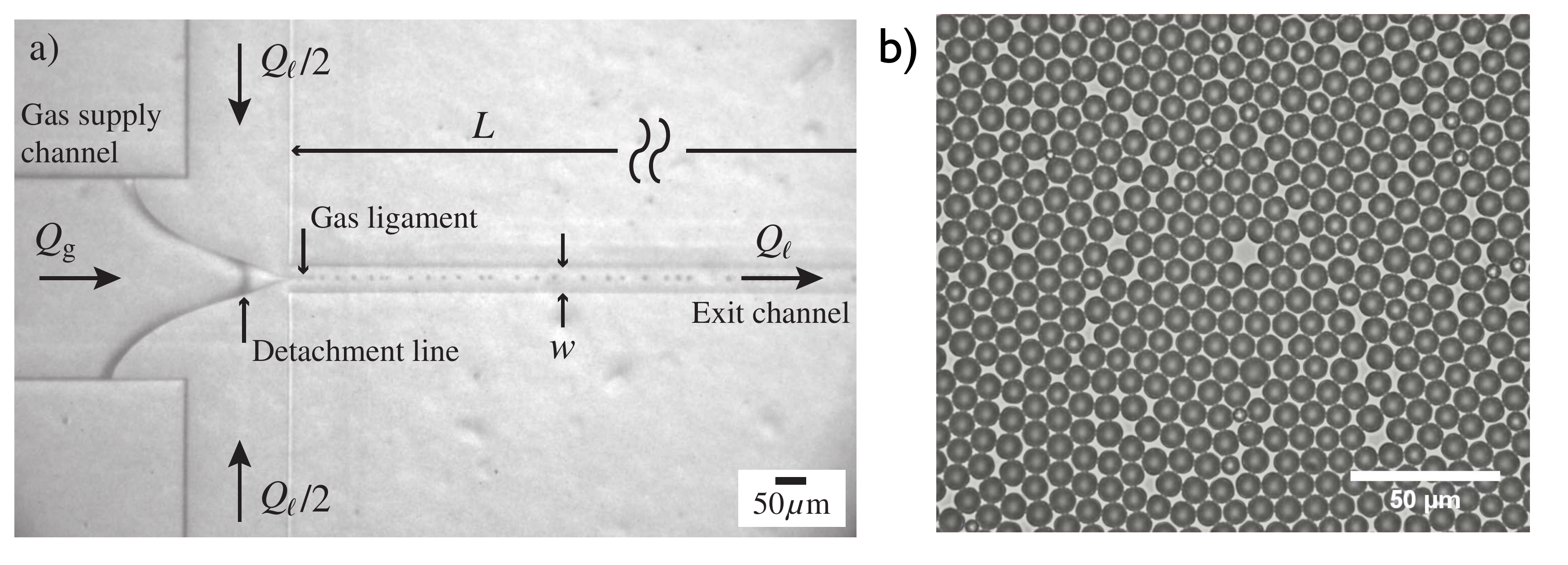}
\caption{
\it
(a) Principles of the co-flow device of ref.\ \cite{castro2011} to produce monodisperse microbubbles. The control parameters are the 
gas flow rate $Q_g$, the liquid flow rate $Q_\ell$, and the geometric parameters, including the thickness (not shown) of this quasi-2D device. 
The scale bar in the lower right is 50 $\mu$m. 
Figure taken from ref.\ \cite{castro2011}.
(b) Monodispersed ultrasound contrast agents produced with such a  co-flow device, employing the principles of ref.\ \cite{castro2011}. 
The scale bar in the lower right  corner is 50 $\mu$m. 
Figure by Wim van Hoeve,
Tide Microfluidics, Enschede. 
}
\label{hoeve}
\end{figure}

Another necessity which arose out of the applications of bubbles as ultrasound contrast agents was
to produce
large numbers of relatively monodisperse and coated microbubbles. The monodispersity is desirable to enhance
the scattering property of the bubbles, which is optimal close to the bubble resonance size given by eq.\ (\ref{rp-res}). 
We achieved this with a so-called co-flow device  originally developed 
by Howard Stone, Dave Weitz and coworkers \cite{anna2003,utada2005,utada2007} and in the case of bubbles by Gordillo and coworkers \cite{gordillo2004}, but now 
operated in a regime in which we could produce particularly small monodisperse 
bubbles in large quantities \cite{castro2011}, see figure \ref{hoeve}. 
Meanwhile this method has been commercialized within  a start-up company, as spin-off from
our Physics of Fluids group. 
 
\begin{figure}[htb]
  \centering
\includegraphics[width=0.88\textwidth]{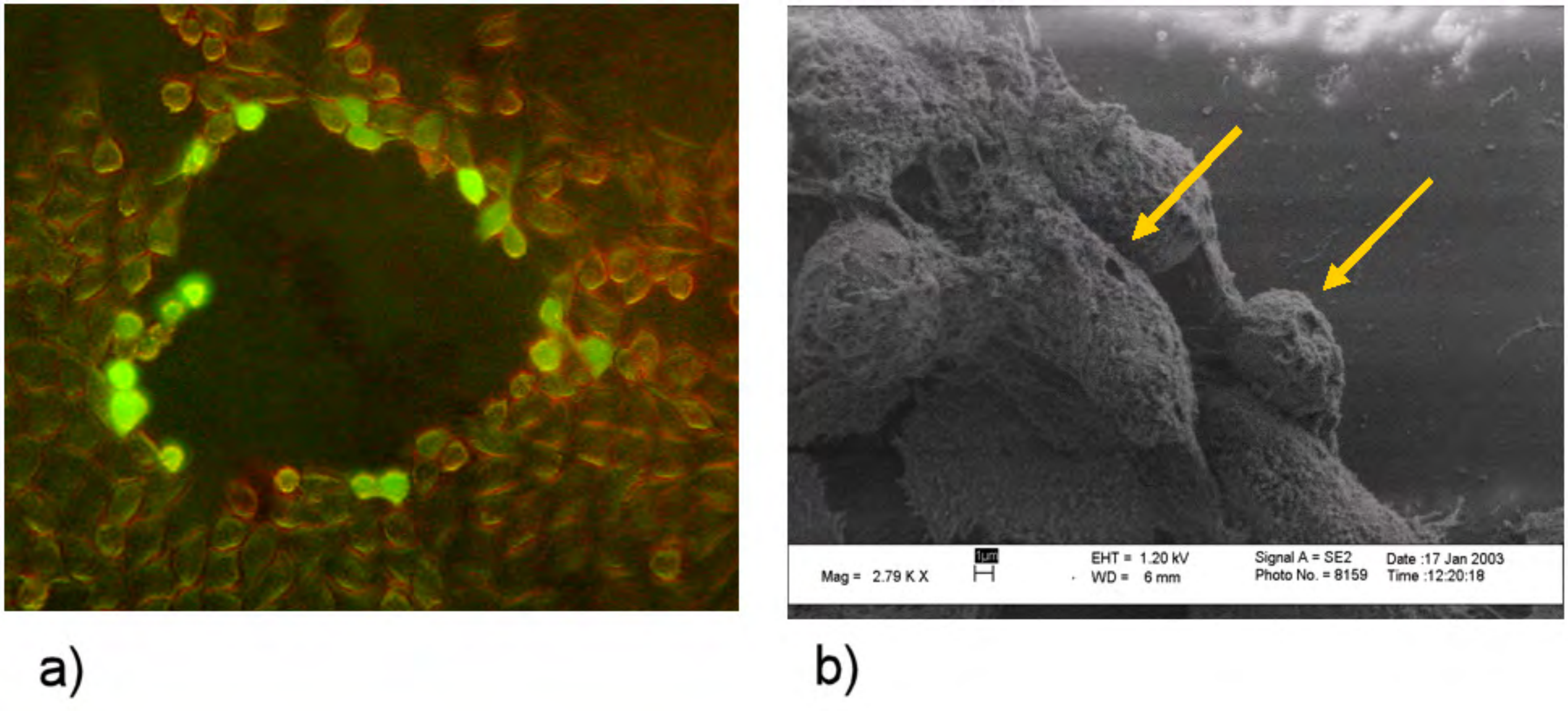}
\caption{\it
(a) HeLa cells (some $\mu$m in size) glued to a glass plate. In the center
of the cell colony a bubble imploded, leading to a cell detachment due
to the induced shear flow. Cells at the edge took up fluorescine, which
is only possible through holes in the cell membrane.
(b) These holes can be visualized through electron microscopy. Figures taken
from ref.\
\cite{ohl06}.}
\label{fig-med}
\end{figure}

The development 
of the Brandaris 128 ultra-high-speed imaging facility also allowed us  to study 
the interaction of ultrasonically driven bubbles with cells. 
This interaction is sometimes spectacular, as seen  in figure \ref{fig-med} 
\cite{ohl06}, which  shows a HeLa cell culture (a commonly used human cell line) grown on a glass
plate, just after a bubble has collapsed close to it. The collapsing
bubble exerts such strong shear forces on the cell that they detach
from the glass plate, or, if they are more remote,  holes in the
cell membrane are induced. These holes, which can close again
after some time, allow drugs or genes to invade the cell. Therefore
ultrasonically driven bubbles can be used for local application of
genes or drugs. This includes employing 
  emulsions of droplets composed of  liquid perfluorocarbons, 
which are  acoustically  activated to undergo a phase change into a bubbly dispersion, a procedure
termed acoustic droplet vaporization
\cite{shpak2014}. 

\begin{figure}[htb]
\centering
\includegraphics[width=0.99\textwidth]{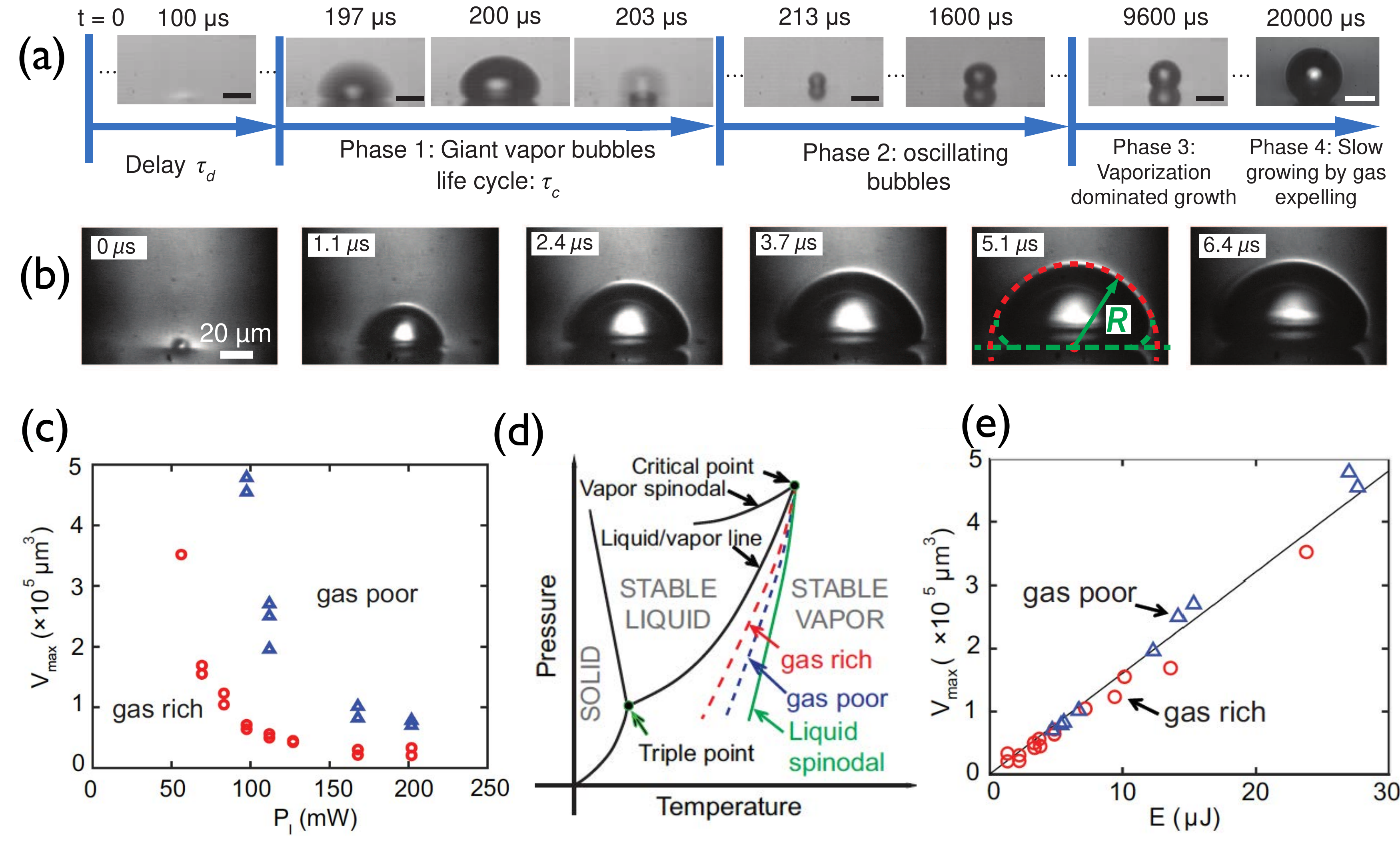}
\caption{\it 
(a) Time sequence of a plasmonic bubble (generated at gold nanoparticles)
 under continuous laser irradiation  in gas-rich water.
  The nucleation and growth dynamics 
   of the plasmonic bubbles as  four phases (see text). The scale bar is 25 $\mu$m.
(b) 
Evolution of the  initial giant plasmonic bubble during its  life cycles in an air-saturated liquid,  
captured at 7.47 million frames per second. 
The  laser power was $P_\ell=185mW$.
(c) Maximum volume $V_{max}$
 of the giant bubble as function of laser power $P_\ell$ in gas-rich water and gas-poor water.
(d) Schematic phase diagram of water. The green solid line is the liquid spinodal line, the theoretical limit of superheat, while the blue and red dashed lines schematically depict the attainable superheat for gas-poor and gas-rich water.
(e) Maximal volume of the giant bubble $V_{max}$ as function of the dumped energy
 $E = P_\ell \tau_d$ in gas-rich and gas-poor water. 
 Both cases show the  identical linear relation, regardless of 
 the delay time $\tau_d$ and the applied laser power $P_\ell$. 
All figures taken from ref.\ \cite{wang2018}.
}
\label{wang}
\end{figure}

I take the opportunity to stress the fundamental 
differences between vapor and gas bubbles, which in detail are elaborated and explained 
by Prosperetti in his recent review on vapor bubbles
\cite{prosperetti2017}. While for gas bubbles  it makes sense to ascribe them an ambient
radius as the gas exchange processes with the environment are slow due to the slow gas diffusion, it does not make sense for vapor bubbles, which are controlled by the much faster heat diffusion and condensation and evaporation. Also the resonance
frequency of vapor 
 bubbles does not scale like the inverse radius as for gas bubbles (equation (\ref{rp-res})), but as 
$\sim 1/R^{2/3}$ \cite{prosperetti2017}. Note that an expanding vapor bubble is not only invaded by evaporating liquid,
but also by  gas dissolved in the liquid, which in the long term crucially determines its dynamics and lifetime,
as we showed for vapor bubbles generated with water-immersed plasmonic nanoparticles \cite{wang2017}.

Such so-called plasmonic microbubbles \cite{fang2013,baffou2014}
indeed also have  potential biomedical applications
 \cite{lapotko2009, emelianov2009, baffou2013, shao2015, liu2014-medical}, again both in diagnosis and therapy 
(next to other potential applications in micro- and nano-manipulation, catalysis,  and 
solar energy harvesting \cite{neumann2013}),
 and understanding and controlling the dynamics of these microbubbles is key to successfully exploit them -- and to recognize potential risks. 
In figure \ref{wang}a we show the life-cycle of such a plasmonic nanobubble, nucleating
in air-saturated water thanks to laser-illumination of  plasmonic gold nanoparticles, each with a
 diameter of about 100 nm. Note the very  different timescales in between the four snapshots.
After some delay time $\tau_d$ after the beginning of the illumination, the bubble explosively grows to giant size (as compared to the size of the nanoparticle), up to a maximum  radius of 80 $\mu m$, and collapses again within $\approx 10  \mu s$ (bubble life phase 1, which we time-resolved with ultra high-speed
imaging in figure \ref{wang}b). The maximum  bubble volume $V_{max}$ remarkably increases with decreasing laser power $P_\ell$, see figure \ref{wang}c, and, also remarkably, decreases with increasing
gas saturation of the water.  

We could  explain \cite{wang2018} these remarkable features, based on 
  the phase diagram of water  (see
figure \ref{wang}d for a sketch) and in particular the lines of attainable superheat therein, which are in between
the line of liquid-vapor coexistence and the liquid spinodal line.
We first measured the delay time $\tau_d$ from the beginning of the illumination up to nucleation, which drastically increases with decreasing laser power, leading to less total dumped energy 
$ E = P_\ell \tau_d$. This  dumped energy $E$ shows a universal linear scaling relation with $V_{max}$, irrespectively of the gas concentration of the surrounding water (figure \ref{wang}e). This finding supports 
the interpretation 
that the initial giant bubble is a pure vapor bubble. In contrast, the delay time does depend on the gas concentration of the water, as gas pockets in the water facilitate an earlier vapor bubble nucleation, which leads to smaller delay times and lower bubble nucleation temperatures (see again the phase diagram of water, figure \ref{wang}d). After the collapse of the initial giant bubbles, first much smaller oscillating bubbles form out of the remaining gas nuclei (bubble life phase 2, up to 
typically 10 ms, see fig.\ \ref{wang}a). Subsequently,
 a vaporization-dominated growth phase takes over and the bubble stabilizes (life phase 3). In the final life phase 4,
  the bubble slowly grows by gas being expelled due to heating of the surrounding.

\begin{figure}[htb]
\centering
\includegraphics[width=0.99\textwidth]{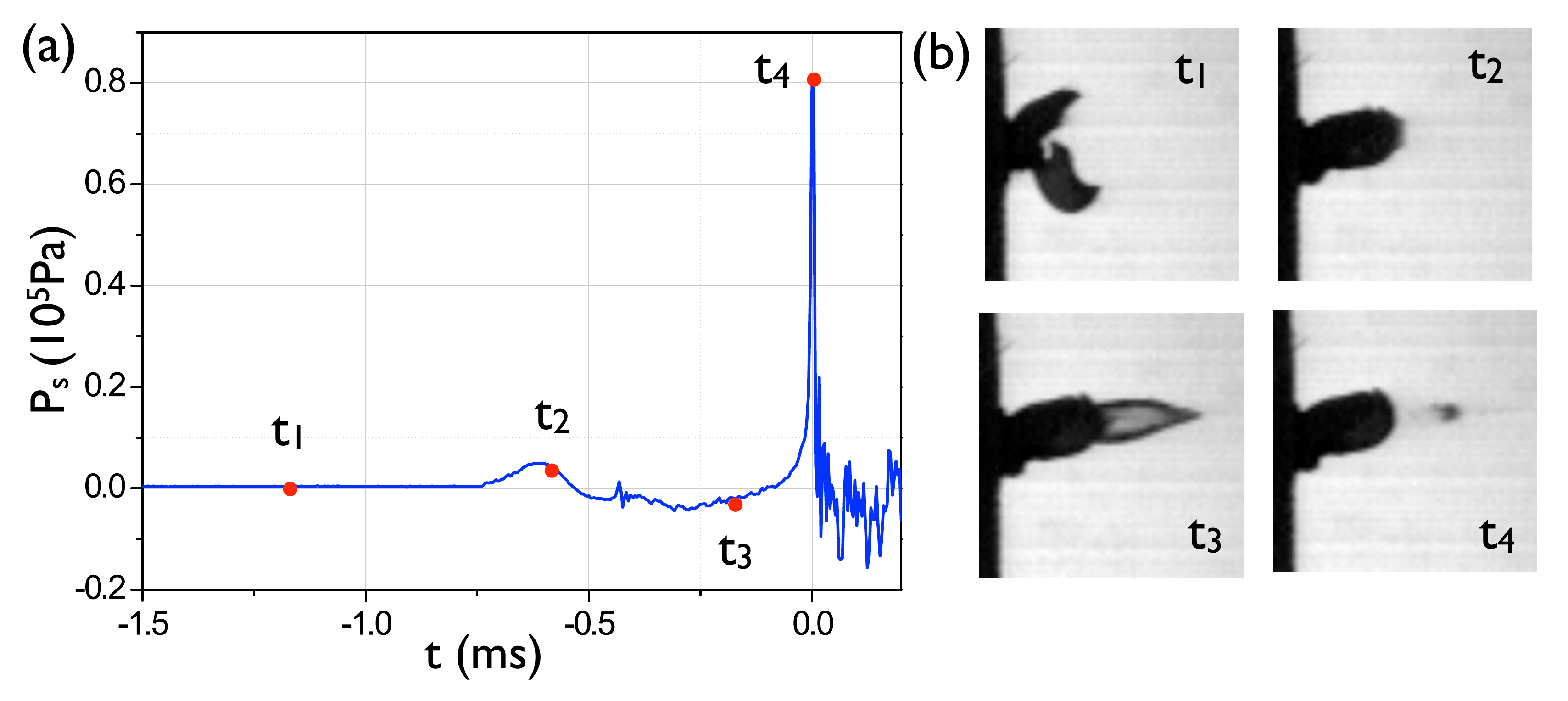}
\caption{\it
(a) 
Sound emission of a snapping shrimp as function of time.  (b) The frames 
 show the closing claw at the four times $t_1$, $t_2$, $t_3$, and $t_4$ indicated in (a): 
 At $t_2$ the claw is
closed, but there is no sound. At $t_3$ the cavitation bubble has nucleated and
grown. The bubble collapse at $t_4$  coincides with the maximum of the sound
emission. The figure adopted  from ref.\
\cite{ver00}.}
\label{fig-shrimp}
\end{figure}

\section{Snapping shrimp and the underwater sound of bubbles}\label{shrimp}

In another outreach effort, in 1999 I gave a colloquium talk 
on single bubble sonoluminescene at the Technical University of Munich (TUM), also addressing
the issue of sound emission from the collapsing bubbles. After the talk,
I met with a TUM zoologist who showed me a signal of the sound emission of 
so-called snapping shrimp, and the same evening I met these animals in the lab. They
 are about 
 5 cm long and live in the tropical ocean. With the
help of a huge claw they can make considerable noise. This animal is
very unpopular with the navy: First, it disturbs underwater
communication between submarines. Second, even worse, hostile
submarines use shrimp colonies to ``acoustically hide'' themselves.

The first obvious question to ask is: 
How does the shrimp make such noise? Zoologists thought that
the sound pulse is caused by mechanical vibration on claw closure.
Knowing the sound emission of a collapsing bubble  from the work on sonoluminescence, 
I  had my  doubts
on this hypothesis. Moreover, I knew the beautiful paper of Prosperetti, Crum, and coworkers
on the underwater noise of rain \cite{pro89}, in which it is 
 shown,  by correlating high-speed imaging and sound detection with a hydrophone, 
 that the noise arising when rain drops fall on a water surface does not originate from the
 impact, but from the oscillations of an entrained bubble. 
 Following their example, 
 we made high-speed movies of the snapping event (the shrimp had to be tickled) 
and correlated them with the corresponding sound track \cite{ver00}.
What we saw was that the shrimp closes its claw so quickly that a
fast water jet develops. High velocities imply low pressure. Just as
in single bubble sonoluminescence this leads to growth of bubbles.
Once the pressure has equilibrated, the bubble collapses, leading to
sound emission (figure \ref{fig-shrimp}) at bubble collapse,
\be
P_s(r,t) = {\rho R\over r} ( 2 \dot R^2 + R \ddot R). 
\label{sound}
\ee
Indeed, the singularity (\ref{sing}) in the RP bubble dynamics is reflected in a singularity $P_s(t) \sim (t_s-t)^{-6/5}$ in the 
sound emission.

The second obvious question to ask is:
Why is the snapping shrimp doing all this? The answer is simple: It
wants to eat! The emitted sound pulse of the collapsing bubble is so
strong that little fish or shrimp get stunned or even killed by it and
are then eaten up by the snapping shrimp. Our explanation
immediately solved another paradox: Why aren't there any snapping
shrimp in the deep ocean, but only in water up to about 50 m depth \cite{everest1948}? The
reason is that the hydrostatic pressure increases with increasing
depth so that eventually the shrimp can no longer generate a
cavitating bubble. Thus it would starve in deeper water.

With our background in single bubble sonoluminescence, 
we could not resist looking into possible light emission from the shrimp-produced collapsing bubbles:
Indeed, there was a faint light emission, a phenomenon we called 
shrimpoluminescence \cite{loh01}. 

 \begin{figure}[h]
\begin{center}
\begin{minipage}{0.6\linewidth}
\includegraphics[width=0.98\textwidth]{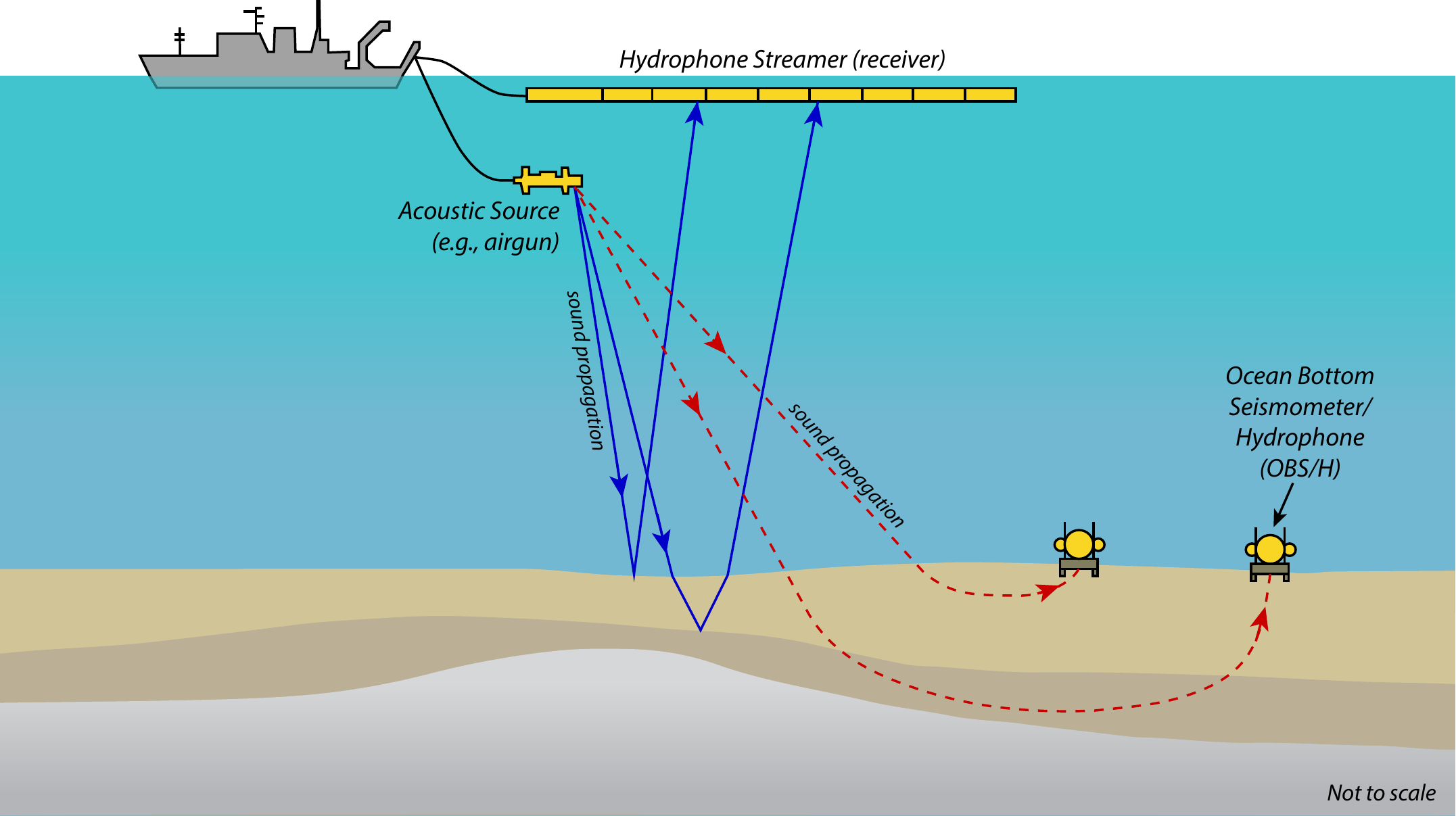} 
\end{minipage}\hfill
\begin{minipage}{0.35\linewidth}
\caption{{  \it 
Principle of acoustic marine geophysical survey.  
Illustration courtesy of National Science Foundation.
    }}
\label{fig-shell-sound}
\end{minipage}
\end{center}
\end{figure}

Sound emission from oscillating or collapsing bubbles is not only relevant for
snapping shrimp, but also on a much larger length scale. Eq.\ (\ref{rp-res}) is not only
valid in the micrometer or millimeter range, but can even be employed as an estimate 
in the meter range, implying that the resonance frequency of a bubble with an ambient radius $R_0 = 1$ m is
3 Hz. Such extremely low frequencies are of great interest,  for example,  in  the oil industry for 
acoustic marine geophysical exploration. The giant bubbles are generated with  so-called
airgun acoustic sources. Their 
 operating principle  is as follows: During the charging
 process, air is put under very high pressure  in a cavity in the airgun. This air is laterally released during the 
 discharging process, leading to a giant air bubble with typical diameters of one meter and beyond.
 The oscillations  of this air bubble leads to sound generation and emission, according to eq.\ (\ref{sound}). 
The sound  gets reflected 
at the ocean-bottom and is 
 detected both with hydrophone streamers and 
with ocean bottom seismometers and hydrophones, see figure \ref{fig-shell-sound}. To extract the relevant information 
from these data, low frequencies around 1 Hz 
are crucial, whereas high frequencies beyond 120 Hz are attenuated in the earth. 
This is the reason why it  is essential to generate low frequencies with the airgun (or airgun clusters)
and thus employ big bubbles.

 Bubbles of this size will no longer be strictly spherical during their whole period of life,
 and this will affect their sound emission behavior, both with respect to intensity and direction.
 The way to calculate the dynamics of such large bubbles or voids from the Navier-Stokes equation
 (or, to be precise, from its  potential flow approximation)
  are 
  boundary integral methods, in the context of fluid dynamics again pioneered by Andrea Prosperetti 
  \cite{pro93d}. 
  With such methods one can optimize the sound emission in the low frequency range.
   This is not restricted to single bubbles, but one can  also analyse bubble clusters
and their collective sound emission, either in 2D by employing some symmetry, or fully 3D. 
  Presently, together with the oil company Shell, who contacted us on this problem, we are
  pursuing such calculations.

\section{Impact on liquids and on ``soft'' sand}\label{impact}

We had started to work with boundary integral methods in the context of impact events.
What had 
 triggered us was again the above mentioned pioneering paper by Prosperetti, Crum and coworkers on  rain 
drops  falling on the ocean \cite{pro89,pro93d}, leading to bubble entrainment and sound emission. Apart 
from the question on sound emission, another  
 important question to ask is: How much air
ends up in the water? The answer is relevant,  for  example, in  climate
models, in which models for the atmosphere are coupled to those of
the ocean. To answer this question, Prosperetti and coworkers had employed boundary integral methods 
\cite{ogu90,pro93d,ogu95}, finding very good agreement with the experimental results. 

In the late 1990s and in the first two decades of this century, the development of digital high-speed cameras has boomed, ever
increasing in frame-rate, resolution,  and storage, and lowering in price considerably. That gave us the opportunity
to look into the impact events and the subsequent void collapse 
in more detail. This line of research was also triggered by single bubble sonoluminescence,
namely to analyze in detail the hydrodynamic singularity at collapse; here not the
spherically symmetry bubble collapse, but the axially symmetric void collapse. 
In fact, mathematically 
the collapse of the void formed  after impact can approximately be described by
a two-dimensional Rayleigh equation
\cite{ogu93,loh04a,bur05,gor05,ber06} analog to the 3D Rayleigh equation (\ref{rp}), which 
has been so successful in describing the collapse of the
sonoluminescing bubble. The 2D version of the inertial part of the Rayleigh equation reads $R \ddot R +   \dot R^2 = 0$, with the singularity solution
$R(t) \propto (t_s - t)^{1/2}$. However, the collapse of a void emerging at impact is not purely 2D and correction terms emerge. Indeed,
experimental studies have found that the exponent of the power law is higher than 1/2
(typical values found are 0.54 -- 0.60) \cite{bur05,tho07,kei06,gor05,ber06,ber09}
 and theoretical studies
have shown that the exponent indeed has a weak dependence on the logarithm of the
remaining collapse time, approximating to 1/2 only asymptotically at the end \cite{gordillo2006,eggers2007,gekle2009}.

\begin{figure}[htb]
\centering
\includegraphics[width=0.98\textwidth]{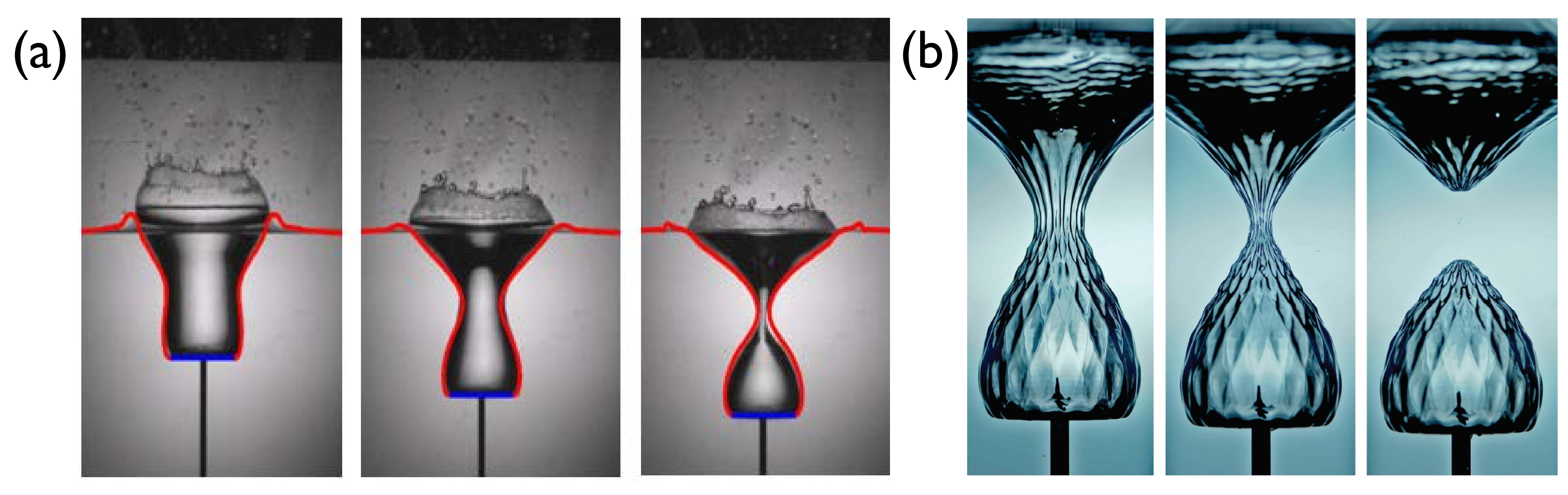}
\caption{
\it
(a) A disk (diameter 6 cm) is pulled through an air-water interface with constant
velocity of 1m/s. The emerging void is collapsing due to the hydrostatic
pressure. The photos are taken 88ms, 115ms, and 131ms after the impact
of the disk, the solid line results from a boundary integral calculation
without any free parameter and shows excellent agreement with the data.
 The pictures are taken from ref.\
\cite{ber06}. 
(b) Here the pulled impact disk (diameter 4 cm) has a small  azimuthal asymmetry with a mode m=20 and 
an amplitude of 4\%. Again, the impact velocity is 1 m/s. The three snapshots show the evolution of the 
shape distortions. 
Taken from ref.\
\cite{enr12}.
}
\label{fig-impact}
\end{figure}

\begin{figure}[htb]
\begin{center}
\begin{minipage}{0.6\linewidth}
\includegraphics[width=0.5\textwidth]{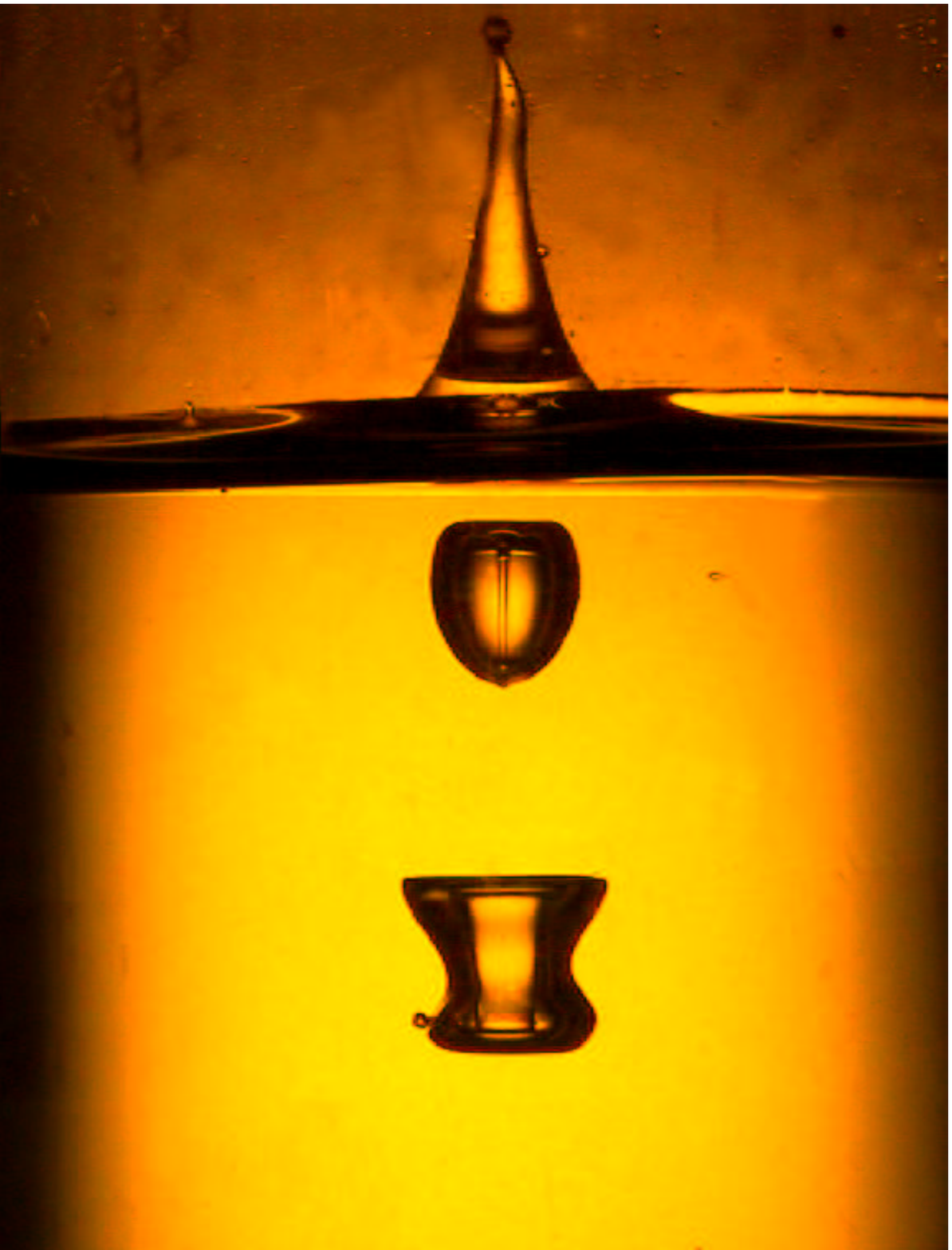}
\end{minipage}\hfill
\begin{minipage}{0.35\linewidth}
\caption{
\it 
A comparable process as in figure \ref{fig-impact} at a later stage: Two jets have developed at
the singularity: One upwards and one downwards into the bubble. This photo
is taken from ref.\
\cite{loh03a}.}
\label{jet}
\end{minipage}
\end{center}
\end{figure}

To analyze these questions in a  controlled way, rather than letting droplets or spheres fall on a water surface,
we 
pulled a disk with controlled velocity $V$ (defining the  Froude number $Fr = V^2 / (gR_{disk})$ as dimensionless control
parameter)
 through the air-water interface and  performed
 high-speed imaging of the void collapse and jet formation \cite{ber06,ber09,gekle2009}.  
  Figure \ref{fig-impact}a shows how the  cavity develops and 
then collapses due to the hydrostatic pressure from the side. At
singularity two jets emerge (figure \ref{jet}): one upwards
straight into the air, the other downwards into the developing
bubble. Just as in 3D the focusing power of the collapsing (sonoluminescing) bubble is converted 
into sound and light emission (and of course heat),  in 2D this focusing power is converted into the jet formation. 
As one can see from figure \ref{fig-impact}a, 
excellent agreement between experiments and the boundary integral simulations can be achieved.
Even the effect of the air flow can be included \cite{gekle2010}, which reverses during the collapse from 
downwards  during void formation to upwards during void collapse, leading to a supersonic air flow out of the closing
void.

The analogy between the 3D bubble collapse and the 2D void collapse goes so far that even 
the shape stability can be analysed in one-to-one analogy, as again first done by Prosperetti
\cite{pro77d}.
Later, we extended this analysis both theoretically and experimentally to find the shape of a collapsing 
non-axisymmetric impact-created air cavity in water
 \cite{enr12}, see figure \ref{fig-impact}b. 

\begin{figure}[htb]
    \centering
\includegraphics[width=0.9\textwidth]{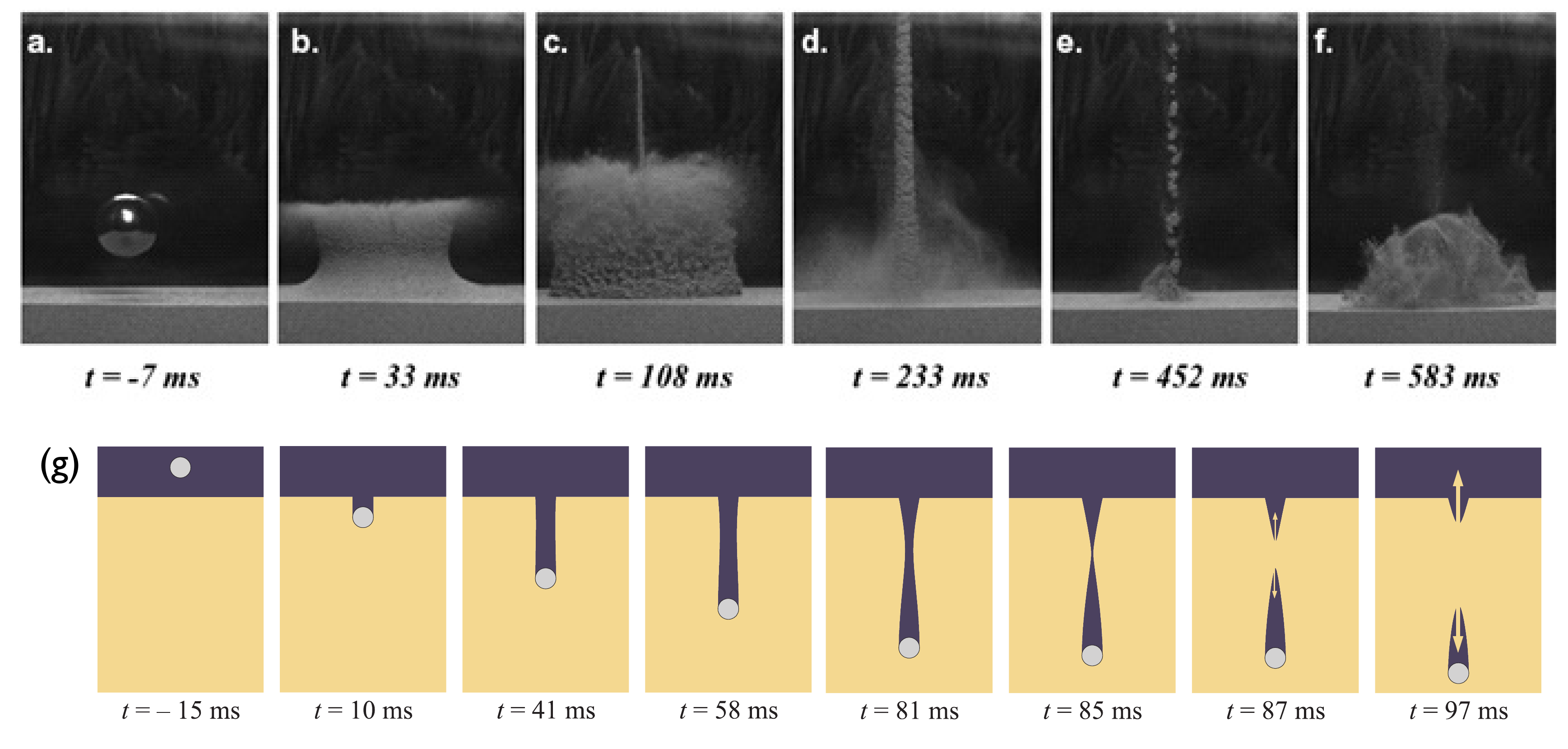}
\caption{
\it (a-f)
Impact (at t=0 s) of a steel ball (diameter 2.5 cm) on soft, decompactified
sand (grain diameter typically 40 $\mu$m). The splash and the jet
emerge, just as in water. The grains in the jet cluster due to their
inelastic collisions. The last frame shows a granular eruption
caused by the rising air bubble. (g) A cross section of the void collapse 
following from the axially symmetric  Rayleigh-type model of ref.\ \cite{loh04a}:
 The void is pressed together by the hydrostatic pressure from the side, leading to a singularity and an upward and downward jet.
 Both  series of images are
taken from
ref.\
\cite{loh04a}.
}
\label{fig-impact-sand}
\end{figure}

To proceed to even larger scale: How comparable is the impact of a
ball on a water surface with that of an asteriod on the surface of a planet? We
downscaled such an astroid impact to lab-scale by fluidizing very
fine sand. Before the impact event, the air-flow is turned off. The
first  reason for the fluidization is to create reproducible
conditions. Second, the fluidization implies that the energy stored
in the ground decreases by  orders of magnitude, corresponding to
the much smaller kinetic energy of the falling ball in the lab as
compared to that of an impacting astroid. The idea is to achieve
similar Froude numbers and Newton numbers 
(the ratio between yield stress of the surface and kinetic energy of the intruder)
as in the geophysical event,
hoping for similar dynamical behavior.

Indeed, the  phenomena of the impact of a ball on such
prepared sand turned  out to be very comparable to those of the impact
on water \cite{loh04a}: First a splash is formed and then a jet
develops (figure \ref{fig-impact-sand}), just as in the water case (figure \ref{jet}). Even
the bubble, which forms in water, again develops and slowly rises, finally, when hitting the sand-air interface, 
causing a granular eruption. It however also turned  out that the air in between the sand grains has a
major role in the emergence and intensity of the jet \cite{cab07,royer2007,homan2014}, namely, when prior to the impact event
(partially)
evacuating the air from the container with the sand, the jet is much less pronounced as the impacting 
 object can intrude  less deep, leading to a weaker ``hydrostatic" collapse. 
An excellent review on impact of objects on granular beds  can be found in ref.\ \cite{meer2017}.

\begin{figure}[htb]
   \centering
\includegraphics[width=0.98\textwidth]{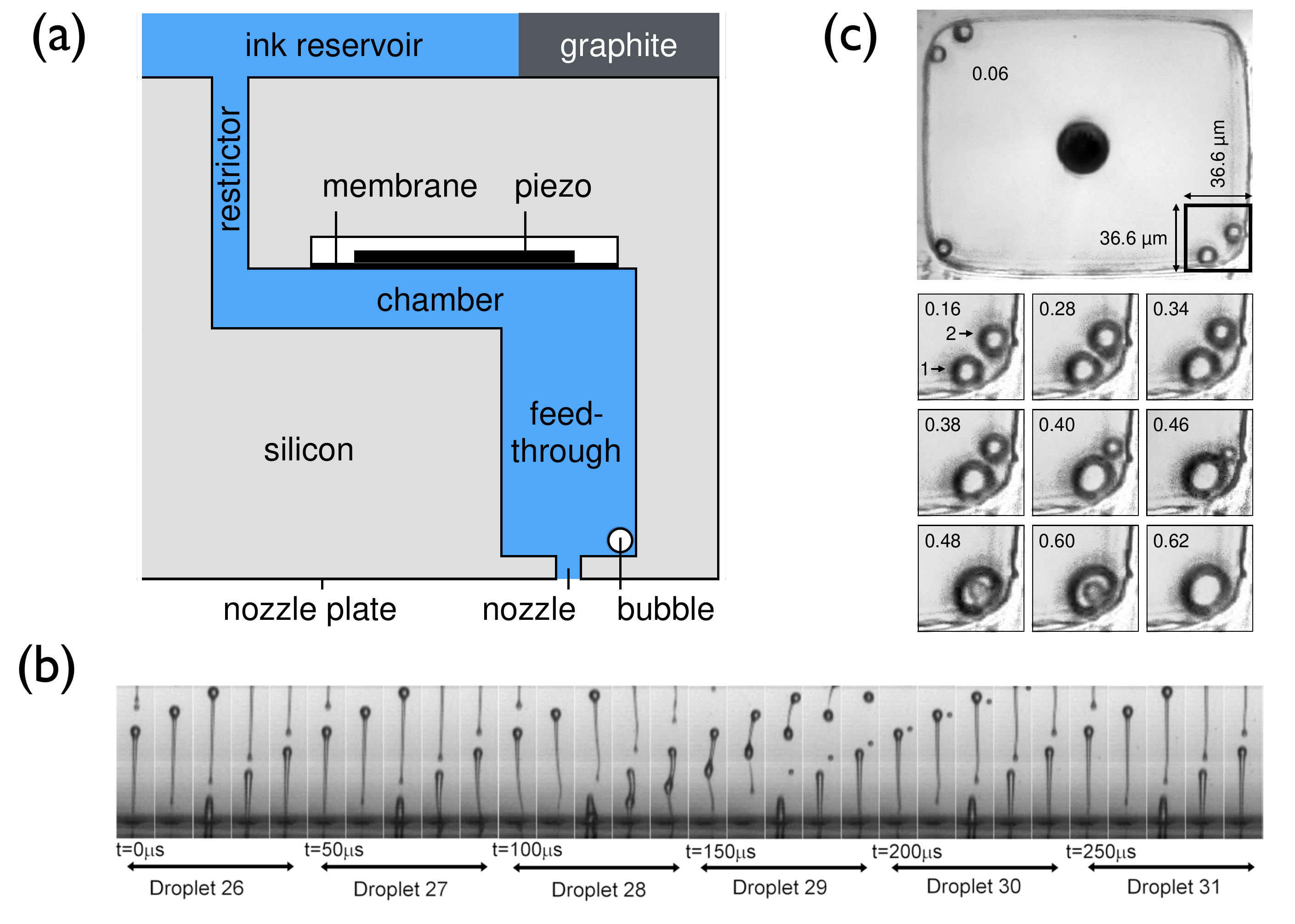}
\caption{\it 
(a) Schematic  of the ink channel (side view) of a MEMS-based drop-on-demand inkjet printer. 
An air bubble has been entrained and pushed to the corner of the ink channel. 
(b) Droplet distortion (droplets 28-29)
caused by a dirt particle around the jetting nozzle. 
Such a dirt particle can lead to bubble entrainment. 
In the shown case here the normal 
droplet formation process recovers and becomes  regular again. 
Figure taken from ref.\ 
\cite{jon06}.
(c) Upper: Infrared bottom
 view through the silicon around the nozzle into the ink channel. In the center the nozzle is seen.
In three of the four corners bubbles got entrained which affect the printing process. The nine smaller images show
the diffusive dynamics  of the bubbles, 
clearly revealing bubble growth (by rectified diffusion) and Ostwald ripening. 
The times are given in seconds. Images taken by Arjan Fraters, Physics of Fluids group, Twente, in collaboration with Oc\'e. 
}
\label{fig-inkjet-distorted}
\end{figure}

\begin{figure}[htb]
   \centering
\includegraphics[width=0.98\textwidth]{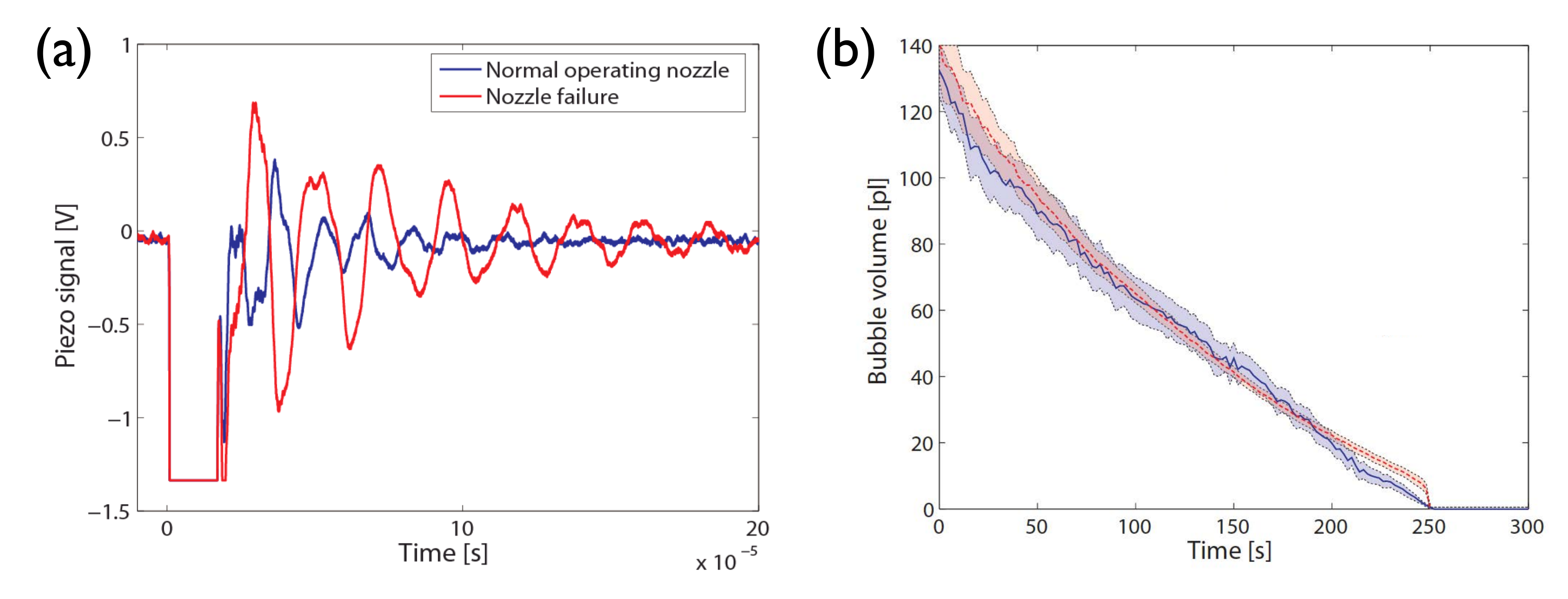}
\caption{\it 
(a)  Acoustic response (as reflected in the piezo-current) of a normally 
 operating nozzle  (blue line)
  and with an entrained air bubble with a volume of $V_b=80 p\ell$  (red)
   close to the nozzle plate. It can be seen  that the volume oscillations of the entrapped bubble modify the piezo-current significantly: the piezo-current amplitude is less damped and the main frequency decreases.
(b) Comparing optically and acoustically measured bubble volume during bubble dissolution (which takes about 250 s here): 
The acoustically measured bubble volume  is shown as  red 
dotted line and the optically measured bubble volume  as blue 
 solid line. The areas around the lines give the error margins in the results.
Figures taken from ref.\ \cite{jeu09}. 
}
\label{inkjet-hear}
\end{figure}

\section{Piezo-acoustic inkjet printing and immersion lithography}\label{inkjet}
Science in the 
university and science in industry often have a difficult time finding each other.
In an effort to facilitate the contact, in 2001 the Dutch science foundation 
organized a get-into-contact event, where I met Hans Reinten from Oc\'e, at the time an independent Dutch company, and since 2010
 a member of the Canon Group. Today, Oc\'e is the leading developer of high-end production printing systems for commercial printing. 
Oc\'e has been  developing  
piezo-acoustic ink-jet printers and obviously various great fluid dynamics challenges come with this,
starting from the flow in the nozzle, to the jetting and droplet process, down to drop impact, drop spreading and drop-paper interaction,
and finally (partial) drop evaporation or solidification. One of the most burning questions 
Oc\'e had in those days was on the fluid dynamics
in the print-head. 

A schematic cross-section through a typical modern MEMS-based 
print-head is seen in figure \ref{fig-inkjet-distorted}a. The piezo-actuator -- similar to the
piezo used to generate single bubble sonoluminescence -- gives short
pulses with -- depending on the printer -- a frequency of around  20 kHz to 100 kHz. 
Each pressure pulse drives out a
little droplet. Altogether, typically there are presently hundreds of nozzles within one
silicon chip and several chips are integrated in one
print-head. A modern inkjet printer has tens of such  print-heads. 

Hans Reinten then  told me that 
unfortunately this very fast and precise printing facility can
break down from time to time: After billions of cycles a distortion
of the droplet formation in a channel can develop. This distortion
either vanishes after a short time (figure \ref{fig-inkjet-distorted}b), or the jetting
process of that channel eventually completely breaks down. The only
solution then is to turn off the piezo and wait for a minute or so,
which of course is extremely annoying for a high-speed printer (though other nozzles might
take over during that time). 
The suspected culprit for 
the problem was a bubble within the inkjet channel. But how does the bubble get there and what is its dynamics? And how to
avoid this trouble?

To solve this  problem, we employed the same method as for the
snapping shrimp \cite{ver00} or the entrained bubble at water droplet impact \cite{pro89}: watch and listen. I.e., we measured the acoustic
response of the channel and combined it with high-speed imaging.
Indeed, we found that the distortion of the droplet 
is correlated
with a modification of the acoustic response of the channel. This
result indeed suggested
 that the distortion originates from a  bubble,
because bubbles modify the acoustical behavior of the channel \cite{jeu08,jeu09}. 
In fact, we were even able to ``hear'' the size of the bubble, see figure \ref{inkjet-hear}. 
But how does the bubble get there? Is it nucleated or entrained at the nozzle? 

By combining  high-speed imaging for the inkjet 
and infrared imaging for the interior of the ink-channel, 
we  succeeded to visualize how a bubble is  entrained at
the nozzle \cite{jon06,jon06b} and what are  its dynamics inside the channel 
(figure \ref{fig-inkjet-distorted}c). 
Here small  dirt particles -- either on the nozzle plate or in the ink channel -- play a crucial role. 
Once a tiny bubble is entrained at the nozzle, 
the acoustical forces pull
it  into the channel. Just as in single bubble
sonoluminescence, the oscillating bubble then grows in the acoustic field
by rectified diffusion. So the knowledge which we had acquired from single bubble sonoluminesence -- namely on 
acoustical forces on a bubble and on rectified diffusion -- was essential in solving this problem. 

Once the bubble has grown by rectified diffusion to considerable size, 
the actuation  pressure pulse to jet a droplet  simply leads to bubble
compression, and not to a pressure increase at the nozzle. Therefore, no
droplets can be jetted any longer. Only after  the acoustic
pressure has been turned off, can the bubble dissolve by diffusion so that 
printing becomes possible again.

The final
goal of course must be  to avoid the entrainment of the bubble or to
immediately get rid of it again, e.g., by applying an acoustical
pulse immediately after the bubble has been detected. For the next
step -- the development of even faster printers with even smaller
droplets -- further fundamental work on the meniscus instability leading to the bubble entrainment 
remains essential.

\begin{figure}[htb]
\begin{center}
\includegraphics[width=0.98\textwidth]{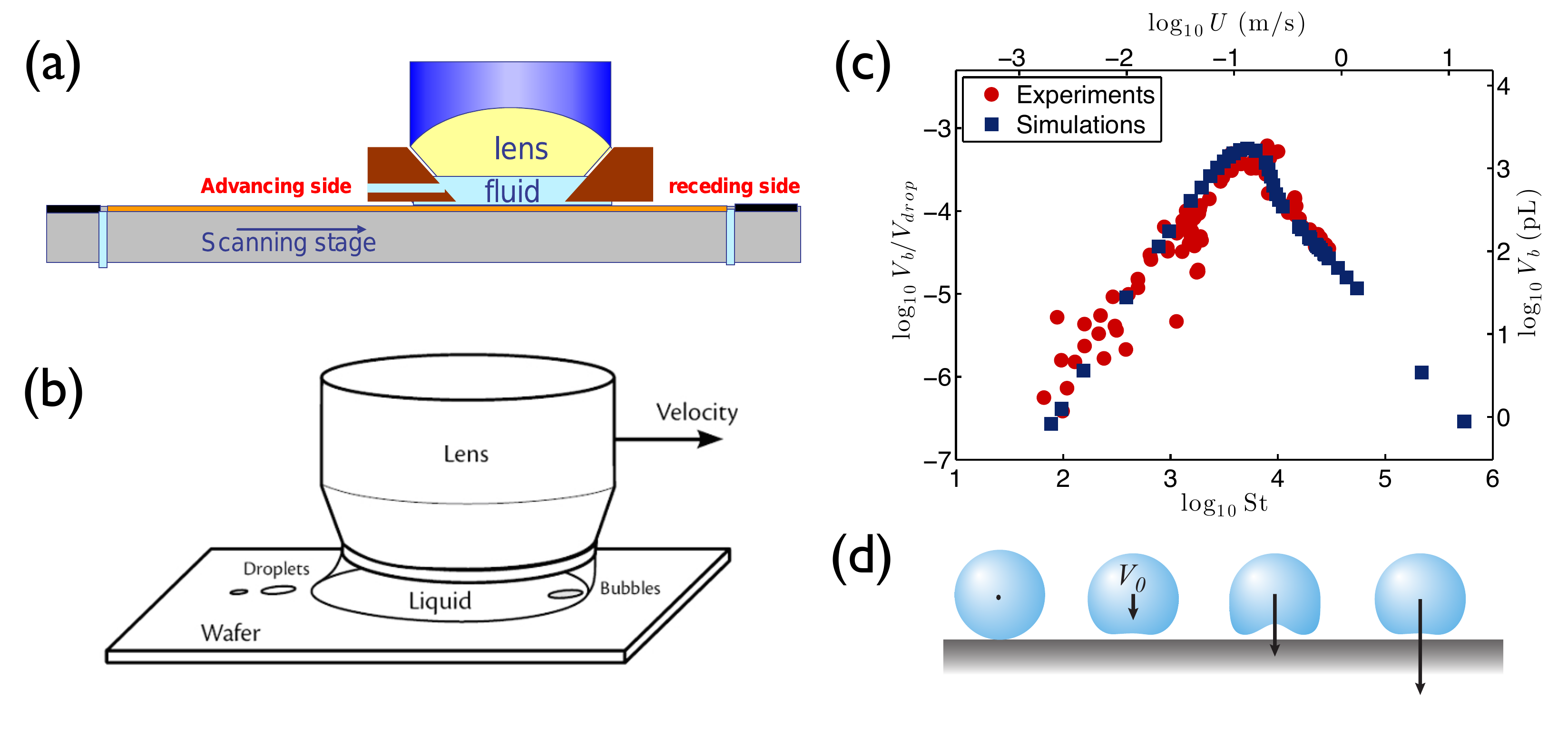}
\caption{
\it
(a) Schematic cross section of an immersion lithographic scanning device. 
The water in between the lens and the wafer reduces the imaging resolution below 40 nm. 
(b) Visualization of the contact line instabilities: 
Bubbles are entrained at the advancing contact line whereas droplets are lost at the receding contact line.
(c) Entrained bubble volume after drop impact versus impact velocity, given both in dimensional units 
(right and upper axis) and dimensionless units (left and lower axis), showing a clear maximum. To the left  capillarity
prevails, while to the right inertia prevails. 
Red circles correspond to experiments with color-interferometry, blue squares to boundary integral simulations
coupled to a viscous lubrication approximation. 
Figure adopted from ref.\ \cite{bou12}. 
(d) Sketch of the droplet impact and bubble entrainment mechanism, with (from left to right) increasing velocity. 
 }
\label{asml} 
\end{center}
\end{figure}

Bubble entrainment also turned out to be crucial for so-called 
immersion lithography, pioneered by  ASML, the world's  leading
supplier  of lithography systems for the semiconductor industry. Once the photolithography of wafers had been at the edge of 
optical resolution,   even smaller structures were realized by introducing
 lithography machines using immersion technology, i.e., lithography under a film of water (with a refractive
  index of 1.33, rather than 1.00 for air) 
 between wafer and optical lens, see figure \ref{asml}a, 
  allowing for smaller structures on the wafer than standard lithography in air. 
 However, the development of immersion lithography machines has been 
  hindered by two  fundamental fluid dynamical problems, namely, the entrainment of bubbles into the water and the loss of water from the film, see fig.\ \ref{asml}b. 
  Both of these contact line instabilities have challenged the further development of immersion lithography, 
  as obviously the entrainment of light-scattering bubbles into the film or the partial loss of the film is 
  unacceptable for the lithography process. 
  These hydrodynamic instabilities set in at  a certain velocity with which the wafers under the lens are
   pulled away. It is thus this
  hydrodynamic instability which sets the production rate of the wafers and therefore 
   the price of the lithographic system. 
  
In the context of this problem, together with ASML we looked at the entrainment of bubbles under droplets impacting on 
a solid substrate, following pioneering work of the Chicago, Harvard, Kaust and other groups
 \cite{xu05,tho05,thoroddsen2008,man09,hic10,kol12} (for recent reviews, see refs.\ \cite{josserand2016,tropea-book2017}).  
 By combining
high-speed color interferometry \cite{vee12}, scaling arguments, and 
numerical simulations with the boundary integral method for the droplet coupled
to a viscous lubrication approximation for the gas flow in the thin and narrowing gap between impacting droplet and substrate, 
we found that there is an optimal
velocity for maximal bubble entrainment
\cite{bou12}: For lower velocities the impacting droplet remains
 more spherical (capillary regime) and for higher velocity (inertial regime)
the drop is smashed against the surface so much that not much air can be entrained, either,
see fig.\ \ref{asml}c. 
Obviously, this work is also of interest 
for the inkjet industry and the coating industry.

\begin{figure}[htb]
\begin{center}
\includegraphics[width=0.98\textwidth]{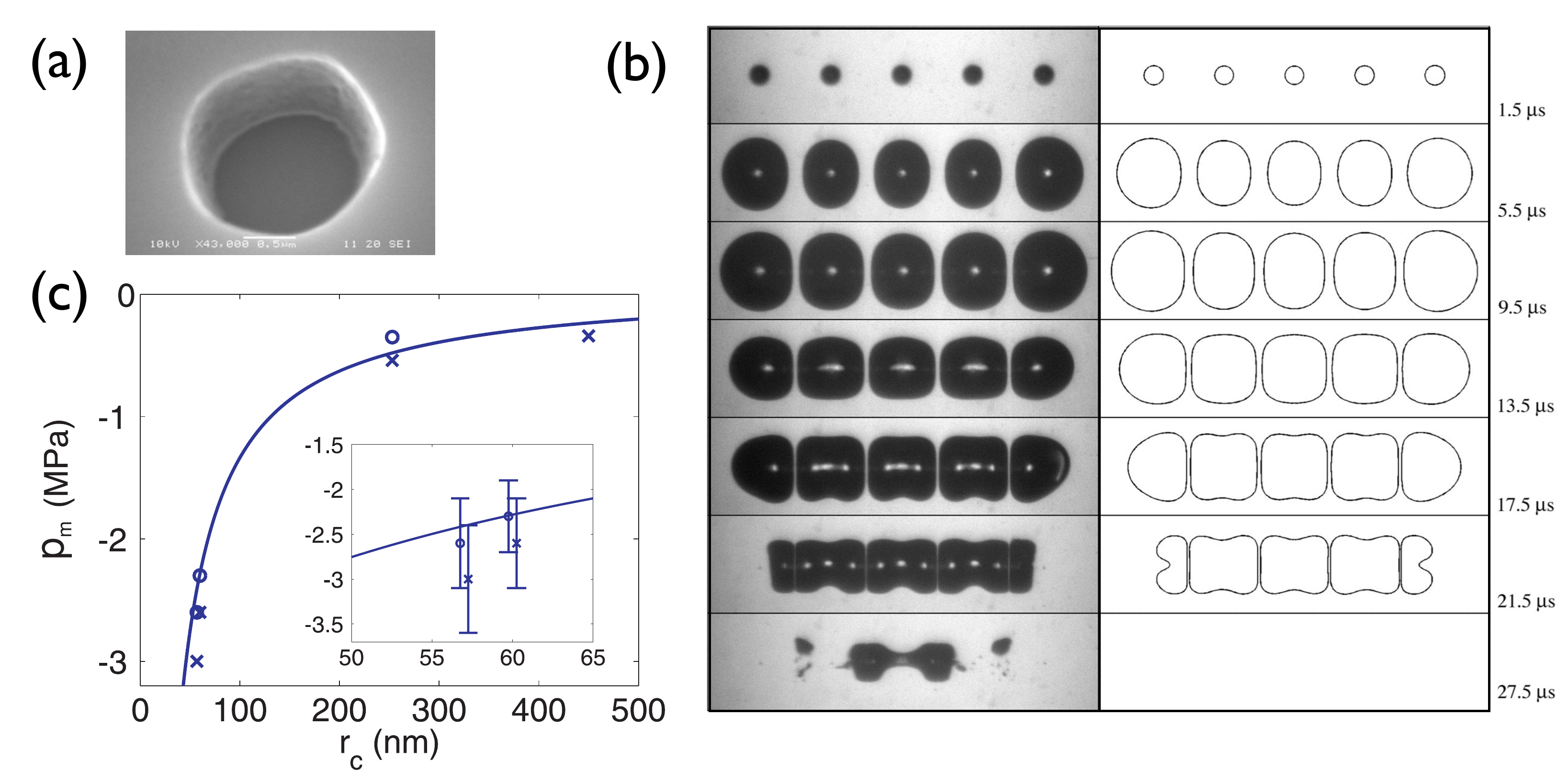}
\caption{
\it
(a)  Electron microscopy image of an individual hydrophobic microcavity (diamter $4\mu m$) etched on a silicon plate acting 
as gas trap. Taken from  ref.\ \cite{bre06b}.
(b)
Comparison between experiment and boundary integral simulation of the cavitation of  5 bubbles in microholes set on a
 line with a distance of $d=200\ \mu$m and a driving pressure of $P_{a}=-1.4$ MPa. 
 One clearly sees the shielding effect for  the inner bubble, collapsing later than the outer ones. 
 Figure taken from ref.\ \cite{bre06b}. 
 (c) Nucleation threshold $p_m$ as function of the pit radius $r_c$ 
 for both theory (line) and experiment (symbols, crosses: nucleation, circles: no nucleation). The inset shows a zoom in with error bars. For visibility overlapping points are shifted $\pm$0.25 nm with respect to each other.
 Figure taken from \cite{bor09}.
 }
\label{bremond} 
\end{center}
\end{figure}

\section{From surface bubbles to sonochemistry \& ultrasonic cleaning} \label{sonochemistry} 

As stated above, single bubble sonoluminescence can be seen as the hydrogen atom of bubble fluid dynamics. But atomic physics did not stop with the understanding of the hydrogen atom but moved ahead to more complicated 
 and interacting atoms, molecules, and condensed matter. So it was also our desire to better understand interacting bubbles,
 but in the  most controlled way, with fixed distance. To achieve this acoustically is difficult, as acoustically driven microbubbles
 either repel or attract each other through the so-called Bjerknes forces of the second kind
 \cite{met97,bre95}, depending on their size and the driving pressure. If trapped by the primary Bjerknes
 forces in different pressure antinodes of the acoustic field,  they are
  too far away (7.5 cm corresponding to the  acoustic
 wavelength for $f =20$ kHz in water) from each other to considerably interact. Therefore we came up with the idea to trap the 
 bubbles through
 pinning forces to  hydrophobic micro-machined 
 micro-cavities which act as gas traps when the substrate is immersed in water
  \cite{bre06,bre06b}, see figure \ref{bremond}a,b. 
  The patterning of the substrate allows us to control the number of  bubbles and the distance between them. Each hemispherical bubble experiences the effect of its mirror image. Correspondingly, an {\it isolated} hemispherical bubble together with its mirror image behaves like a free spherical bubble, i.e., its dynamics is well described by the Rayleigh-Plesset equation (\ref{rp}).
 By putting the micro-cavities close to each other,
  we could study {\it  interacting} microbubbles, either in a row (i.e., in 1D) as in figure \ref{bremond}b, where we could compare the results
 with boundary integral simulations (i.e., under the potential flow approximation)
 with axial symmetry, or in 2D with bubbles arranged on a surface in any order.

 In fact, by varying the diameter of the microhole down to 100 nm and less, we could quantitatively test  \cite{bor09} the crevice model of bubble
 nucleation 
 \cite{harvey1944,apfel1970,cru79,atchley1989}. Figure \ref{bremond}c compares the nucleation threshold calculated from the crevice model 
 \cite{atchley1989} with the experimental data, finding good agreement. 
 We also used such hydrophobic micro-machined pits  as artificial crevices for bubble nucleation  to achieve higher sonochemical yields at ultrasound powers that would otherwise not produce a significant chemical effect \cite{rivas2010} and for ultrasonic cleaning purposes \cite{riv12},
 in both cases making use of the energy focusing power of the collapsing bubbles. 
 Out of this activity another spin-off company emerged from our group. 
 Finally, with such pits  we enhanced the heat flux in thermal convection by vapor-bubble nucleation \cite{guzman2016}.

\section{From surface nanobubbles to catalysis and electrolysis}\label{nb}
\begin{figure}[htb]
\begin{center}
\textnormal{(a)}\includegraphics[width=0.42\textwidth]{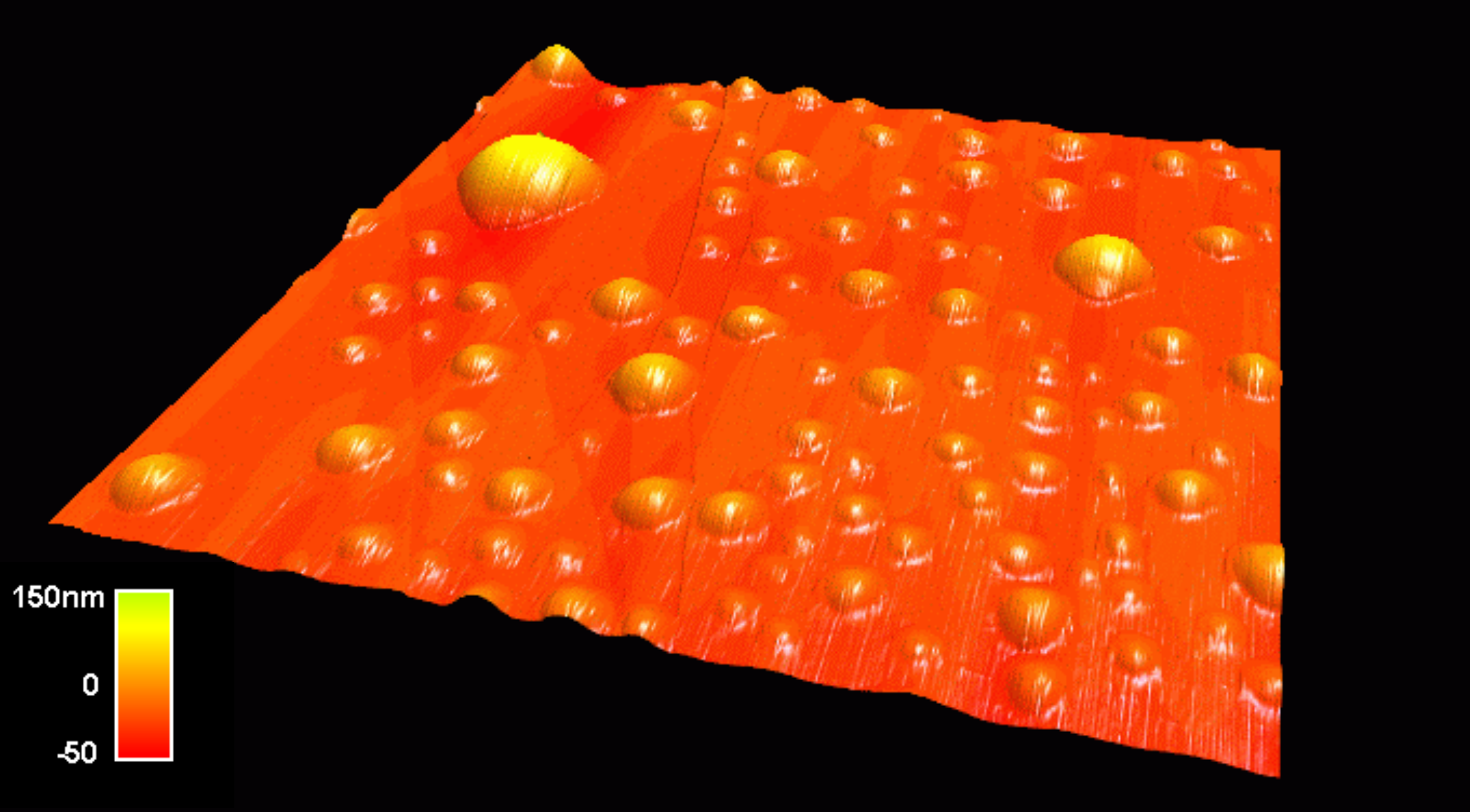}
\textnormal{(b)}\includegraphics[width=0.42\textwidth]{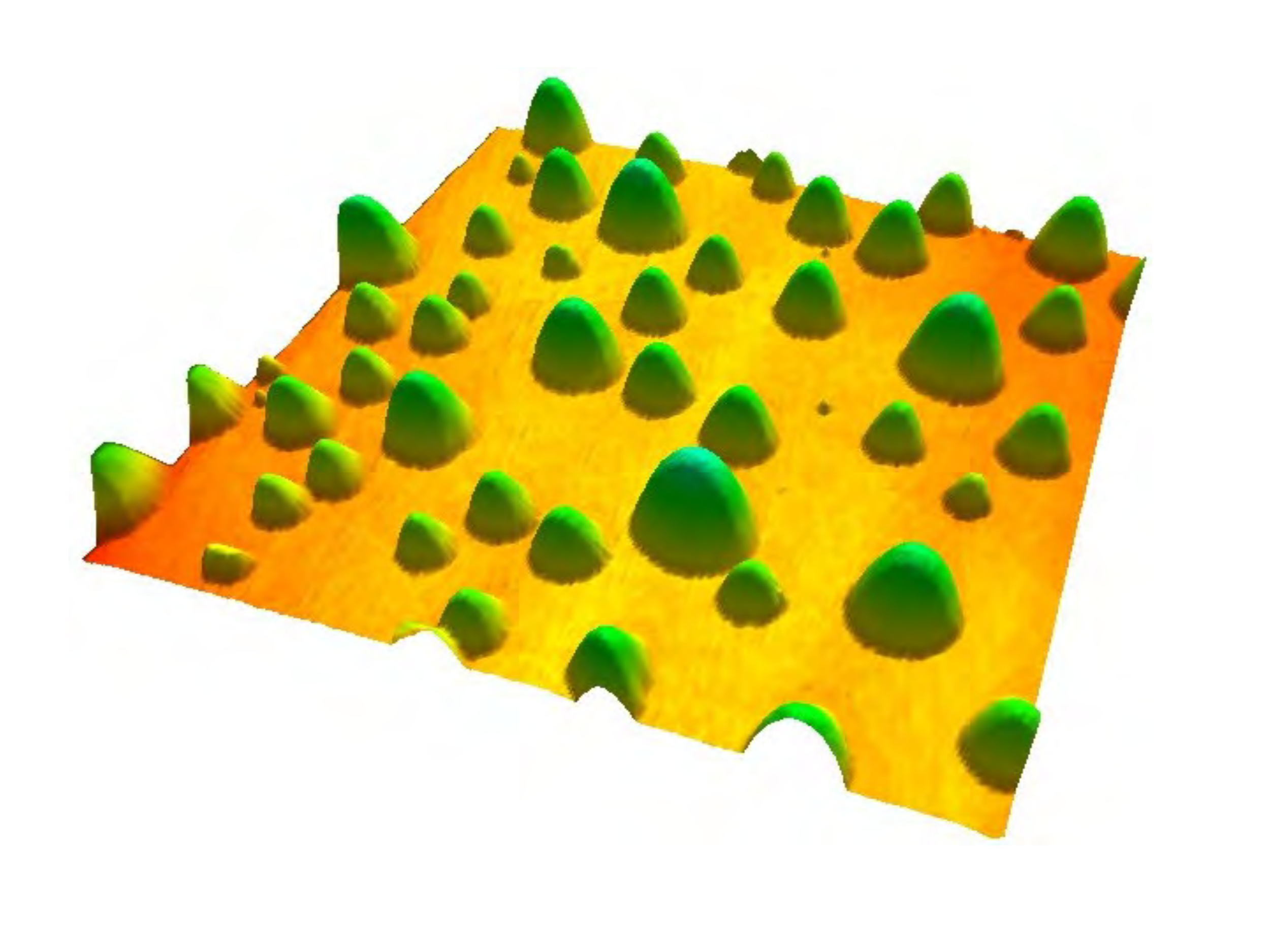}
\caption{
\it (a)
AFM image ($4\times 4 \mu m^2$) of a surface nanobubble on a HOPG surface, obtained through the solvent
exchange process.  
(b) AFM image ($30\times 30 \mu m^2$) of surface nanodroplets on a hydrophobically coated
Si surface, also obtained through the solvent exchange process. The color code goes from 
 0 (red) to  800 nm (green). Figures taken from our recent review article 
 on surface nanobubbles and surface nanodroplets  \cite{lohse2015rmp}. 
  }
\label{image-nb} 
\end{center}
\end{figure}

Being interested in tiny surface bubbles, the so-called surface nanobubbles caught my attention, 
which  from about 2000 on 
 have  been found   in atomic force microscopy (AFM) images  of water-immersed, preferentially hydrophobic substrates \cite{lou2000,ishida2000,tyrrell2002}, see figure \ref{image-nb}a.
One of course immediately wonders why such surface nanobubbles are stable. Because of the diverging Laplace pressure
$p_{Laplace} = 2\sigma / R$,
 where $\sigma$ is the surface tension and $R$ the radius of curvature,
  tiny bubbles should dissolve
immediately: In water the gas pressure inside a bubble with R = 10 nm  is  $p_{gas} = P_0 + p_{Laplace} \approx 145$ atm. 
With Henry's law this translates to a gas concentration at the edge of the bubble $c_R = p_{gas} c_s/P_0$   which is 145 
 times larger than the saturation concentration
$c_s$ and therefore to a large concentration gradient away from the bubble, leading to very fast dissolution. This even holds
in the case of
(slight) gas oversaturation, i.e., as long as  the gas oversaturation 
\be 
\zeta = { c_\infty \over c_s } -1 \label{zeta} 
\ee
is not too large. 
In the  above mentioned 
classical paper by Epstein and Plesset \cite{epstein1950}, known to us from our work on single bubble sonoluminescence,   the dissolution time of such a bubble had been calculated 
from the diffusion equation and the corresponding boundary conditions.
 The result for the typical dissolution timescale is 
$ \tau_{EP} = R_0^2 \rho_g / (2 c_s D)$,   
where $D$ is the diffusion constant and $\rho_g$ the  gas density. For the bubble with R = 10 nm  and the material constants
for water,
 one indeed gets $\tau_{EP} \approx 3 ~\mu$s, i.e., such nanobubbles should dissolve basically immediately.

The surface nanobubbles in the AFM image
figure \ref{image-nb}a were generated  with  the so-called solvent exchange 
process, which had been pioneered by Xuehua Zhang 
\cite{zhang2004langmuir,zhang2007nanodroplet}. 
Here, a gas-saturated liquid having high gas solubility (e.g., ethanol) is replaced by another liquid 
with  lower gas solubility, e.g. with water. This leads to a local gas supersaturation 
$\zeta > 0$  and therefore to the nucleation of bubbles. 
Macroscopically, we know this  effect of everyday life: When  we fill a water glass with 
cold tap water and leave it for a while in a warm room, small air bubbles form on the inside of the glass. 
The reason is that in general 
 tap water is oversaturated with air and 
 gases in cold water dissolve much better than in warm water. If the tap water in the glass slowly warms up 
 to  room temperature, the gas solubility is reduced and bubbles nucleate on the edge of the glass. 
 Depending on the size of the glass, the bubbles  last about four days, as anyone can easily try.
 
First it was speculated that the stability of surface nanobubbles is due to surfactants \cite{ducker2009}, which however for various reasons (explained in 
ref.\ \cite{lohse2015rmp}) could be ruled out. Moreover, with the help of fluorescence lifetime imaging microscopy  it could be shown
\cite{hain2016}, 
that the objects observed with AFM are indeed air bubbles, and not nanodroplets  of a contaminating liquid. In addition to the 
 remarkable stability of the nanobubbles, another paradox was that their contact angle (measured on the bubble side) was not  Young's angle, as known from macroscopic measurements, but much smaller  \cite{song2011,walczyk2014b}.  
 
When there is such a large gap between experiment and theory, numerical simulations often help. 
We therefore performed MD (molecular-dynamics) simulations of surface nanobubbles \cite{weijs2012prl}, and at least those adhered to the theoretical expectation: they dissolved in microseconds.

\begin{figure}
\begin{center}
\includegraphics[width=0.90\textwidth]{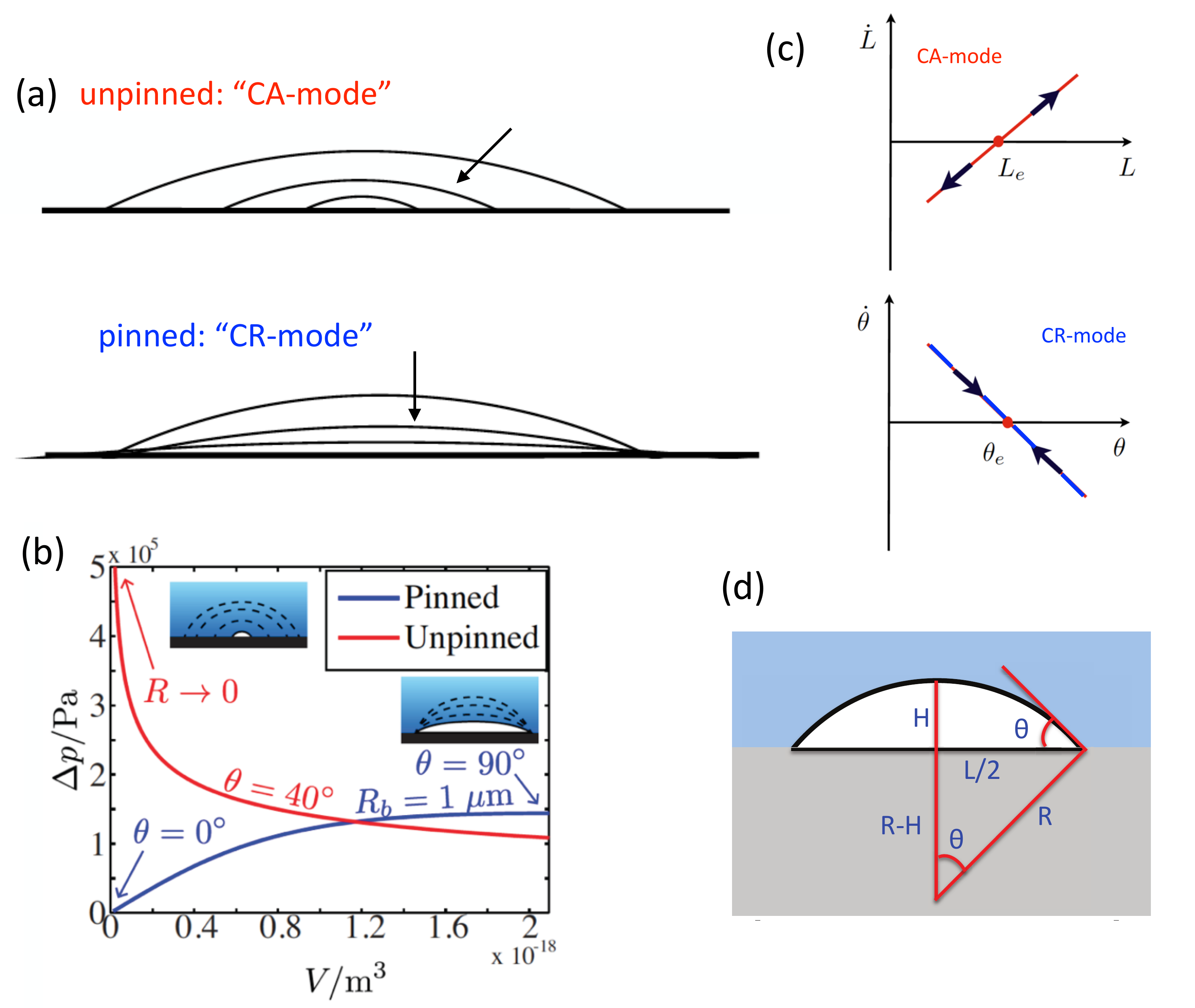}
\caption{\it 
(a) The nanobubbles can dissolve in two different ``modes'' (the time direction is indicated by the arrow): 
In the CA mode (constant contact angle, above), the contact angle is constant. As a result, the radius of curvature becomes smaller and smaller, which leads to the divergence in the Laplace pressure, as can be seen in (b) (red curve), where we show
$\Delta p = p_{Laplace}$ 
 as a function of the bubble volume V. In the 
 CR mode (constant contact radius, bottom), the contact radius is constant and the contact angle becomes smaller. As a result, the radius of curvature increases in the course of the dissolution process and the divergence in Laplace pressure does not occur (blue curve in (b)). 
 In (c),  we show a sketch of the phase space for the stable equilibrium, which results in ``pinning'' (CR-mode, bottom), 
 with the equilibrium contact angle $\theta_e$, given by equation (\ref{theta_e})), 
 and  a sketch of the phase space for unstable equilibrium without pinning (CA-mode), in which the surface bubble either shrinks or grows.
  In (d) the used notation is introduced: $L$ is the lateral extent of the bubble at the substrate (contact diameter), 
  $H$ is the bubble height,  $\theta$  is the contact angle at the gas side,  and R the radius of curvature. 
}
\label{pinned} 
\end{center}
\end{figure}

The key to solving this paradox came from the experiments of Xuehua Zhang and coworkers 
\cite{zhang2013langmuir} who observed that surface nanobubbles exposed to gas-undersaturated water 
($\zeta < 0$)  dissolve slowly, but (initially) not by reducing their lateral size, but by reducing their contact angle (on the gas side), 
see Fig.\ \ref{pinned}a: The three-phase contact line remains pinned. 
This dissolution mode is called ``CR-mode'', standing for constant contact radius, in contrast
to the so-called ``CA-mode'', standing for constant contact angle 
\cite{picknett1977,cazabat2010}.
Pinning dramatically changes the 
dissolution scenario: the Laplace pressure $p_{Laplace} = 2  \sigma/ R$ now 
 no longer diverges, but approaches zero, see Figure \ref{pinned}b. Thus, when the bubble dissolves, no large internal pressure can build up and thus no 
  concentration gradient from the outside of the bubble to the predetermined concentration level 
 $c_\infty$ far away from the bubble: The bubble becomes stable. 
 
  The reason for the pinning lies in the unavoidable surface inhomogeneities of 
  geometric and/or chemical nature. These are also relevant in the above mentioned 
  daily life phenomenon of bubble formation in a glass with cold tap water which warms up or 
  when  we pour soda water into a
  glass: In both cases 
   bubbles nucleate out of oversaturated water on such inhomogeneities. Macroscopically, 
  the surface inhomogeneities lead
  to contact line hysteresis, as studied extensively by de Gennes and co-workers in the 1990s
\cite{joanny1984,gennes1985}.

We could generalize the classical Epstein-Plesset calculation \cite{epstein1950} to calculate the diffusive dynamics of 
pinned surface bubbles \cite{lohse2015}. 
Here the key idea was to adopt the quasistatic calculation of 
Popov \cite{popov2005} for the  so-called ``coffee stain problem''  \cite{deegan1997,marin2011}, 
which is on the evaporation of a liquid drop on a plain substrate. This is not surprising, because both processes are controlled by  diffusion
outside of the drop / bubble: in the evaporating droplet  case, of diffusion of water vapor in air, and in the surface nanobubble case, of   air into water.
For bubbles with constant contact diameter $L$  (i.e., in the CR-mode), 
the result of this adopted (quasistatic) calculation reads \cite{lohse2015}
\be
{d\theta \over dt } = - {4 D\over L^2}  { c_s  \over \rho_g}  ( 1+ \cos \theta )^2 f(\theta ) 
\left[ 
{L_c \over L }  \sin \theta
-\zeta 
\right],  
\label{ode-for-theta}
\ee
with a positive definite   $f(\theta ) $ given in ref.\ \cite{popov2005}
and a critical lateral extension 
 $L_c = 4\sigma / P_0 \approx 2.84 \mu m$ for air bubbles in water with  1 atm ambient pressure. 
From eq.\ (\ref{ode-for-theta}) it  immediately follows that for gas undersaturation 
  $-1  \le \zeta < 0$ no stable surface nanobubbles can exist, as then 
  the right-hand side of eq.\ (\ref{ode-for-theta}) is always negative: the bubble
  dissolves down to $\theta = 0$. For gas oversaturation $\zeta>0$, however, a {\it stable} equilibrium with the equilibrium contact
  angle \cite{lohse2015}
  \be
\sin \theta_{e} = 
 \zeta {L\over L_c} 
\label{theta_e}
\ee
can exist. That the equilibrium indeed is stable is seen from the phase space figure \ref{pinned}c, bottom. 
In this stable equilibrium, 
Laplace pressure (causing gas flux out of the bubble) 
and gas overpressure (causing gas influx) are in balance.

\begin{figure}[htb]
\includegraphics[width=0.9\textwidth]{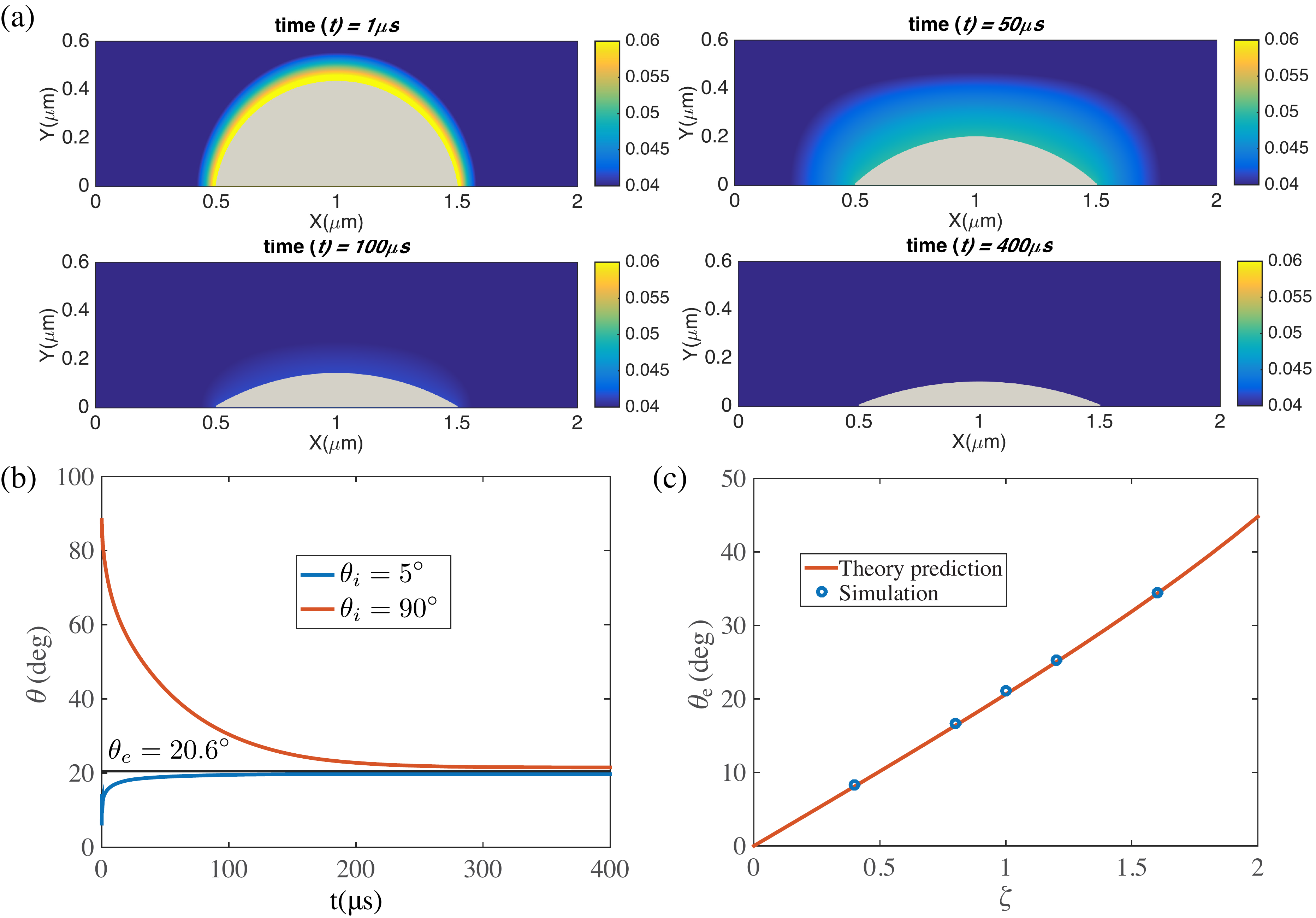}
	\caption{ \it Results from a numerical  simulation (using the finite difference method)
	of the diffusion equation, with the shrinking or growing pinned surface
	bubbles coupled with the immersed boundary method: 
		(a) Snapshots of the diffusive dynamics of a pinned surface nanobubbles growing towards its equilibrium state.
		The color code represents the gas concentration field. Here $L = 1$ $\mu$m and $\zeta = 1$. 
		(b) Time evolution $\theta(t) $ of the contact angle growing or shrinking towards its equilibrium value $\theta_e$ given
		by Eq.\ (\ref{theta_e}). Two cases with different initial contact angles $\theta_i$ are shown. Here,
		$L = 1$ $\mu$m and $\zeta = 1$. 
		(c) Equilibrium contact angle $\theta_e$ for various gas concentrations $\zeta$. The straight line is the prediction eq.\ (\ref{theta_e}),
		giving perfect agreement. Again, $L = 1$ $\mu$m. In the simulations here, the domain size is $6\mu$m$\times$$3\mu$m$\times$$6\mu$m. 		Figures taken from ref.\ \cite{zhu2018}.
				}
   \label{stable}
\end{figure}

\begin{figure}[htb]
\begin{center}
\includegraphics[width=0.9\textwidth]{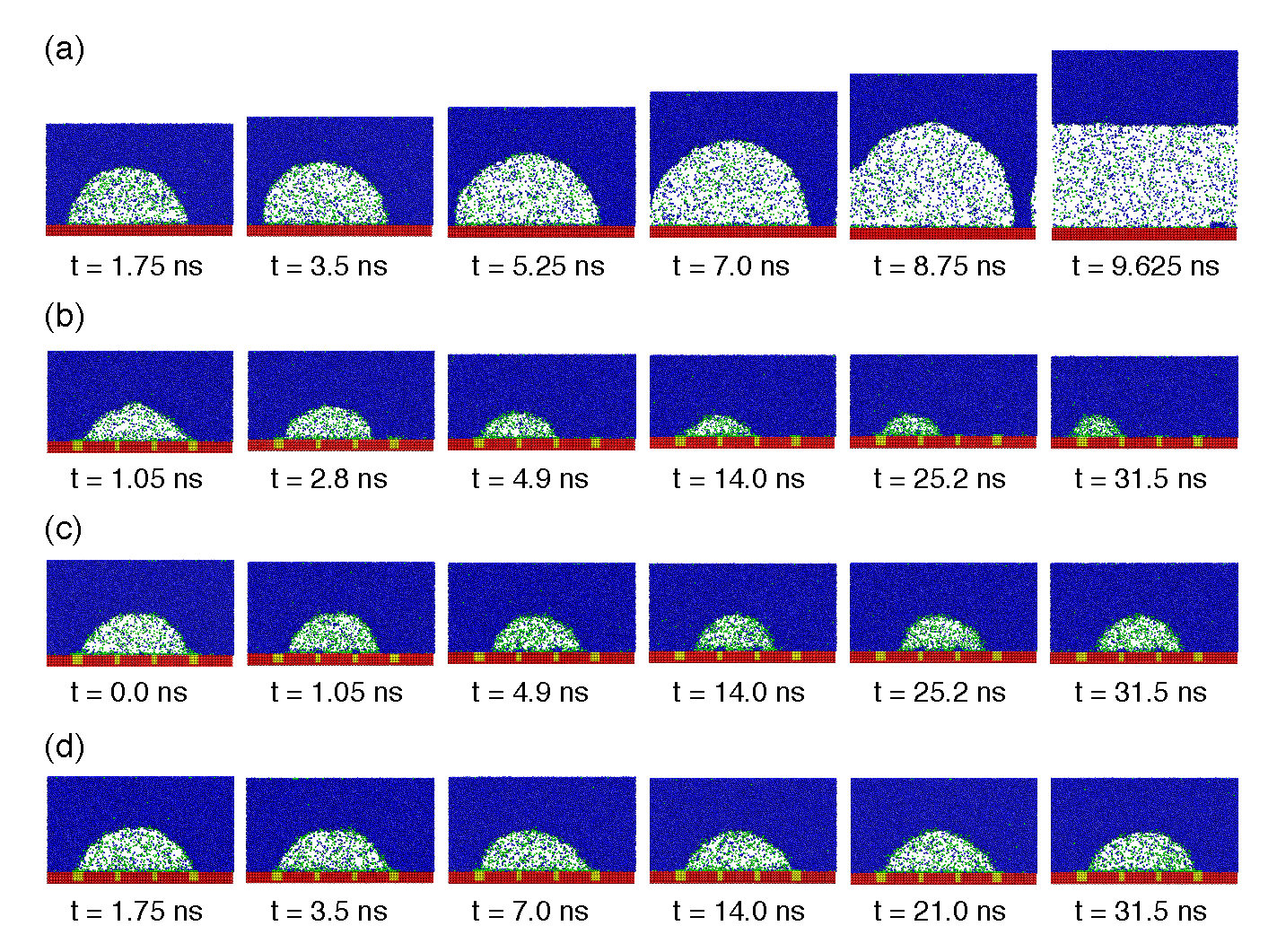}
\caption{
\it   
Time evolution of a surface nanobubble in MD simulations: 
 (a) Without chemical heterogeneities and gas-oversaturated liquid ($\zeta > 0$) the bubble grows.
 (b) With hydrophobic chemical heterogeneities and gas-undersaturated liquid ($\zeta < 0$) the bubble shrinks. 
 (c) With hydrophilic chemical heterogeneities and gas-oversaturated liquid ($\zeta > 0$) the pinning force
 for nanobubble stabilization is not sufficient. 
 (d) Only with hydrophobic chemical heterogeneities and gas-oversaturated liquid ($\zeta > 0$) 
 one does get a stable surface nanobubble. Figure taken from \cite{maheshwari2016b}.
}
\label{md-stability} 
\end{center}
\end{figure}

From equations (\ref{ode-for-theta}) and (\ref{theta_e}) we also see that for too  large oversaturation $\zeta > L_c/L $ there is no stable
equilibrium and the surface bubble keeps on growing so that it will finally detach. The condition $L < L_c/\zeta$ is 
the reason that there can be stable surface bubbles only on a microscopic scale. The stabilization mechanism does not work on a macroscopic scale, because then the Laplace pressure is too weak and cannot compensate for the gas overpressure from the outside. We also see that on the microscopic scale, the radius of curvature of a surface bubble is not given by  Young's equation 
but by the relationship (\ref{theta_e}). 
We note that 
eq.\ (\ref{theta_e}) and the stability of the equilibrium have 
 also been 
 confirmed both by numerical simulations of the full diffusion equation \cite{zhu2018}, employing immersed boundary methods for the 
growing or shrinking bubble, see figure \ref{stable}, and in addition also 
 by MD simulations --  but now with built-in pinning \cite{maheshwari2016b}, see figure \ref{md-stability}.

As we see,  the size $L$ of the pinning site  and the oversaturation $\zeta>0$  
determine the stability of the surface nanobubbles. But how is the oversaturation $\zeta$ determined? 
When the liquid container with  the surface nanobubbles on some substrate 
is closed and in equilibrium, $\zeta$ remains constant and hence the equilibrium
 contact angle $\theta_e$, Eq. (\ref{theta_e}). In an open vessel, on the other hand, an initial gas oversaturation
$\zeta>0$  will not last long due to diffusive processes with the outside world \cite{weijs2013prl}. If the distance to the outside world is 
$\ell$, the typical diffusive time scale is 
$\tau_{outer}
 \sim
\ell^2/D$. For  $\ell = 1 $ cm we get  $\tau_{outer} \approx 14 $ hours and 
for $\ell = 3 $ cm  $\tau_{outer} \approx 5 $ days:
This is exactly the time scale that we also observe for  the dissolution process of  air 
bubbles which nucleate at the edge of the above mentioned  glass filled with cold, gas-supersaturated 
tap water when put into a warmer room.

\begin{figure}
\begin{center}
\includegraphics[width=0.99\textwidth,angle=-0]{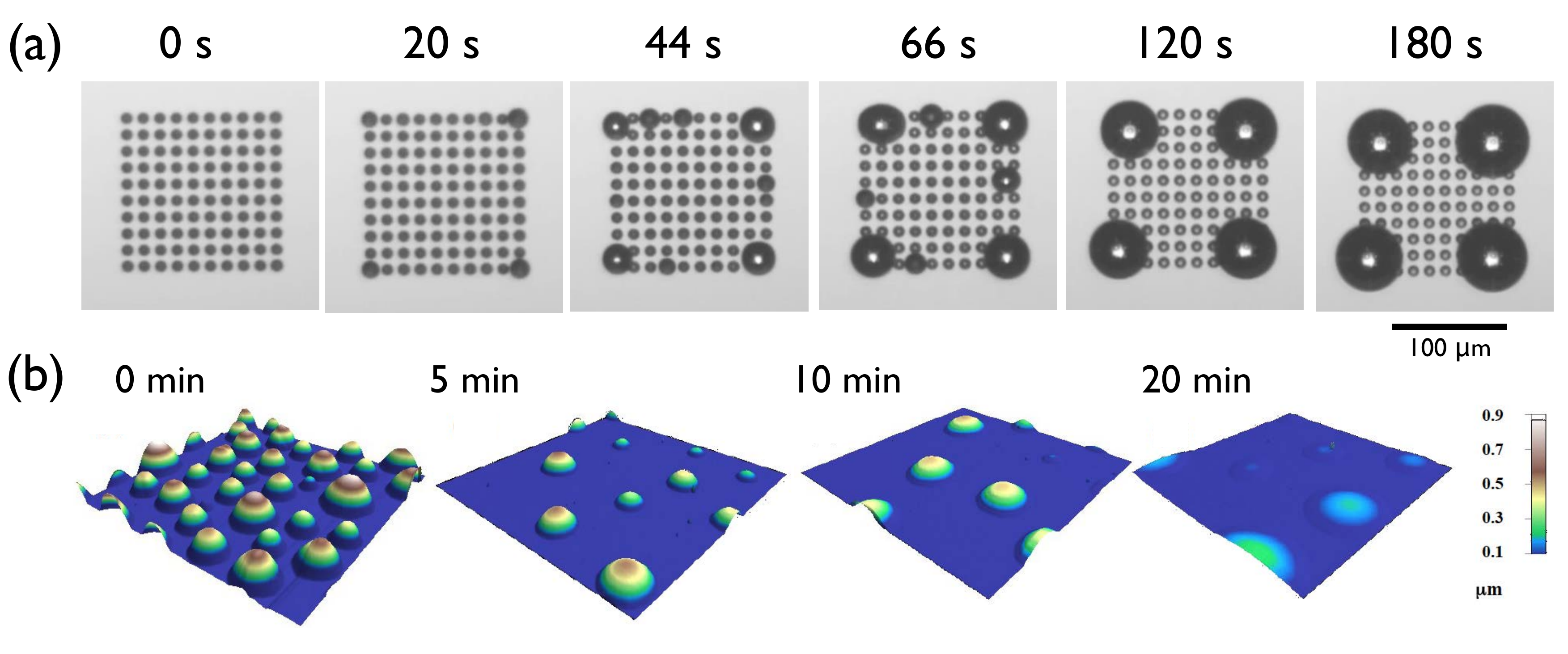}
\caption{\it 
(a) Controlled microbubble formation and growth on hydrophobic surfaces decorated with 
micro-machined 
micropits (with diameter of 10 $\mu$m and  edge-to-edge 
distance of 13 $\mu$m) by solvent exchange:
The microchannel  was first filled with air-saturated water which was then 
replaced by ethanol solution under a flow rates of 100 $\mu$L/min.
Figure taken from ref.\ \cite{peng2018}.
(b)
Dissolution process of an  ensembles  of  HDODA   nanodroplets with a typical contact diameter of 
 $L = 10 - 30 \mu m$.  The color scale corresponds to a height of  $0 \mu m $ (blue) to $0.9 \mu m$ (red) of the droplets.  
The dissolution process was stopped by  polymerisation of the  HDODA by  UV-radiation, so that a visualization of the droplets
by 
AFM became possible, a trick introduced in ref.\ \cite{zhang2012softmatter}. Figure  
taken from  \cite{zhang2015}. 
}
\label{xh} 
\end{center}
\end{figure}

Just as single bubble sonoluminescence can be seen as the 
hydrogen atom of inertial bubble dynamics \cite{bre02},
a single surface nanobubble can be seen as the hydrogen atom of diffusive bubble dynamics. Its properties 
-- in particular its  on first sight surprising stability and its small contact angle -- are meanwhile
reasonably  well understood, see our review article ref.\  \cite{lohse2015rmp}. As a next step,
 we have  moved towards diffusively interacting surface nanobubbles.
 Just as for inertial bubbles (section \ref{sonochemistry} and figure \ref{bremond}b), one can best study also the diffusive behavior of 
 surface bubbles
  by fixing their 
 distance by offering ``weak spots'' on a hydrophobic surface, namely by 
  micro-machining the surface: During the solvent exchange, the bubbles will nucleate in the cavities and grow  \cite{peng2016-coll}. 
In 
fig.\ \ref{xh}a we show snapshots of this growth process over a period of three minutes.  Note the very different timescale as
compared to the inertial bubble dynamics of figure \ref{bremond}b, where the whole series of snapshots is less than 25 $\mu$s.

For  interacting surface nanobubbles, 
according to above sketched theory \cite{lohse2015} one would expect that in equilibrium all surface nanobubbles would 
have the same radius of curvature $R_e = L_c / (2\zeta )$. On first sight, one may  expect that this 
equilibrium is unstable due to Ostwald ripening of the bubbles \cite{voorhees1985}: Small bubbles shrink due to their larger Laplace
pressure and neighboring larger ones grow. However, it  turns out 
that again it is pinning  which stabilizes these neighboring bubbles against Ostwald ripening \cite{dollet2016,maheshwari2018}.

As stated above, a central idea of solving the surface nanobubble paradox  originated from the 
analogy to the pinned coffee stain \cite{deegan1997}: The first   problem is controlled by diffusion of 
air in liquid, the latter  one by diffusion of vapor in air. In between these two cases is the one of a liquid surface 
nanodroplet (see figure \ref{image-nb}b)
 in a 
sparsely miscible host liquid, for which  the stability (with pinning, or lack thereof, without pinning) 
is given by the same equations and mechanisms \cite{lohse2015rmp}. 

Not only the dissolution processes of such surface nanodroplets (see figure \ref{xh}b)
are completely analogous to those of surface nanobubbles, but also the nucleation and growth processes. In particular, it is again the solvent-exchange process that allows surface nanodroplets to be generated in a controlled manner \cite{zhang2012softmatter,zhang2015pnas}:
 Now  a high solubility liquid saturated with a certain substance is replaced by another with less solubility. 
 The emerging oversaturation leads to droplet nucleation. 
 
As an example, consider the Greek liquor ``Ouzo'' (alternatively Pastis (France) or 
Sambuca (Italy) or 
Raki (Turkey)): Ouzo consists of alcohol, water and anise oil, which is very soluble in alcohol. When served, 
the ouzo is diluted with water. As a result, the alcohol concentration and thus the solubility of the oil
 decreases: Oil droplets then nucleate and  grow in the supersaturated solution.  
 This leads to the known milky appearance
 of the ouzo-water mixture. In the presence of a hydrophobic surface, nucleation naturally takes
 place at this surface and  surface nanodroplets are formed in complete analogy to the surface nanobubbles. With the help of the control techniques of microfluidics, one can study very precisely the dependencies of the nucleation process on the control parameters such as flow velocity, channel thickness, oil concentration, size and shape of the pinning sites etc. and understand them theoretically and quantitatively  \cite{zhang2015pnas,bao2016,bao2018}.

 The technological relevance of 
  surface bubbles and droplets and the solvent-exchange process can hardly be underestimated. In our review article \cite{lohse2015rmp}
  we have dedicated a whole chapter to it; here I restrict myself to some particularly relevant and beautiful examples.

Liquid-liquid extraction -- the transfer of a substance from one solvent to another --  is one of the key processes of chemical technology and analysis. Since the work of 
Nobel Prize winner Fritz Pregl \cite{pregl1917}, efforts have been made to further miniaturize processes for chemical analysis and chromotography. The need to do this has  increased in recent years  \cite{jain2011}. 
First, it is important to be able to demonstrate ever smaller traces of chemical substances, be it in a medical context or for environmental reasons. 
Moreover,
this proof must be
 fast and small samples should be sufficient, also to come to a ``greener''
  chemical technology. One way to achieve this  is single drop microextraction, where a 
   substance A dissolved in water slowly dissolves in a drop of an organic solvent with a much higher solubility for A (e.g.\
    carbon tetrachloride) held in the water. The drop is then extracted and analyzed. However, this 
     process generally takes a long time and the efficiency is low.

These limitations were elegantly overcome with the so-called disperse liquid-liquid microextraction (DLLME) technique 
\cite{rezaee2006a,rezaee2010,zgola2011} invented just 12 years ago. The technique involves a third liquid
 that is  miscible with water and with carbon tetrachloride, e.g.,  ethanol. If a mixture of carbon tetrachloride and ethanol is now added to the water containing the substance A to be extracted, 
 instantaneously many  nano- and microdroplets of carbon tetrachloride nucleate,  due to the enhanced water concentrations these
 molecules experience and the resulting lower solubility. 
 With their very large total surface area, these nano- and microdroplets are very efficient to extract  the substance A. 
 As a final step, the carbon tetrachloride microdroplets  are centrifuged off.
 The  highly enriched substance A can then easily be detected 
 by conventional methods. So far, DLLME is mostly optimized by trial-and-error; however, to achieve further control and
 progress, from my point of view
 it is very promising to apply 
 the well-controlled methods of microfluidics and fluid dynamics to this process
 \cite{lohse2016ff}. 

A good understanding and control of the diffusive processes in and around microdroplets and microbubbles
 may 
  even save lives: in open-heart surgery, the blood stream is 
   decoupled from the heart and maintained by a heart-lung machine 
    \cite{wagner2014} and at the same  time the blood is 
    cooled down. In the cold blood   more air  dissolves
    than when it is  warm. However,
    warming up the patient after surgery can be seen as 
     a solvent-exchange process that can result in the nucleation of (surface) nano- and microbubbles that 
     may  significantly affect the blood supply locally. Here, too, it is essential and possibly life-saving to well understand bubble nucleation
       in diffusive processes.

  \begin{figure}[htb]
  \centering
\includegraphics[width=16cm]{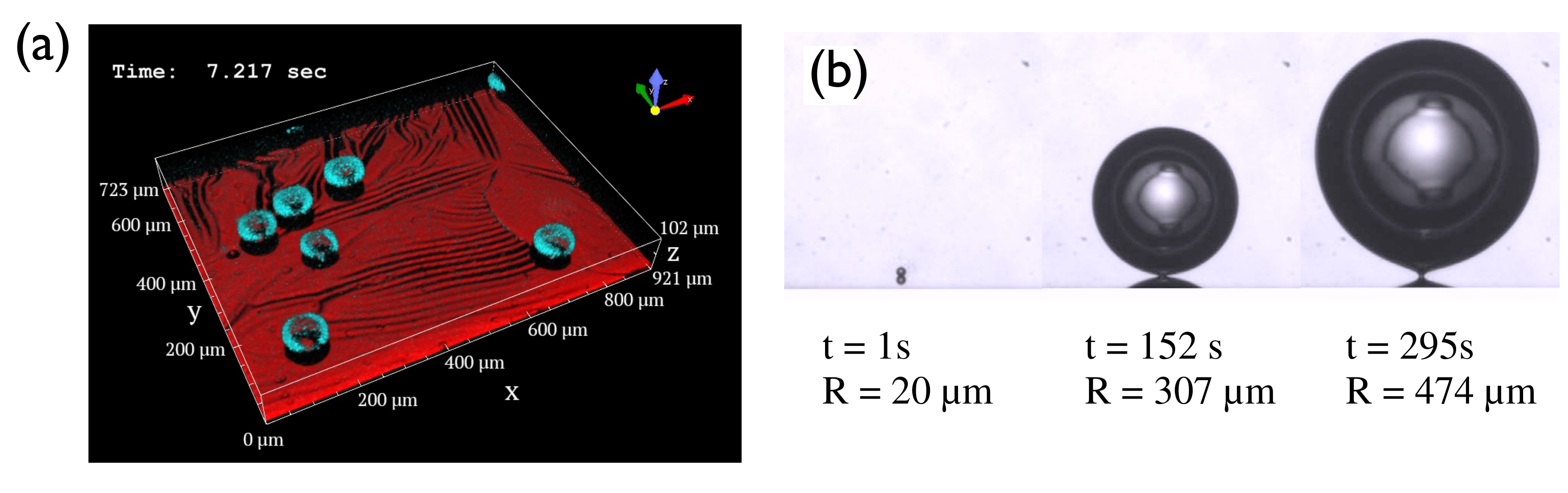}
\caption{\it 
(a) Confocal images of surface oxygen bubbles in a $H_2O_2$-solution (0.1\% in water), growing on a (catalytic) platinum
surface, 7.2 s after exposure of the platinum with the peroxide. Image made by Pengyu Lv, Physics of Fluids, University of Twente. 
(b) Three snapshots of the same surface bubble, pinned at a micropit, and growing out of a CO$_2$-oversaturated
solution with a (pressure controlled) oversaturation of $\zeta = 0.18$. Figure taken from ref.\ \cite{enriquez2014}. 
}
\label{diff-vs-cat}
\end{figure}

Further examples are electrolysis and catalysis: In both processes, the forming surface nano- and microbubbles considerably impair the efficiency,
as they cover the electrode or catalyst.  How does one  control the growth of the bubbles 
and how to get rid of them? 
We have taken 
 some first steps to answer these questions \cite{yang2009,lv2017,linde2017,moreno2018}.  
Figure \ref{diff-vs-cat}a shows catalytically grown oxygen 
bubbles on a platinum surface  immersed in a hydrogen peroxide (H$_2$O$_2$) solution. The platinum
acts as a catalyst which triggers the decay of the peroxide to oxygen (evolving in bubbles) and water. 
For such bubbles evolving out of 
chemical reactions the relevant dimensionless number coming into play is 
the Damk\"ohler number Da, which is the ratio between the chemical reaction rate and the 
(diffusive or convective) mass transport
rate. If Da is very large, the chemical reactions is   controlled by the latter. 

\begin{figure}[htb]
  \centering
\includegraphics[width=16cm]{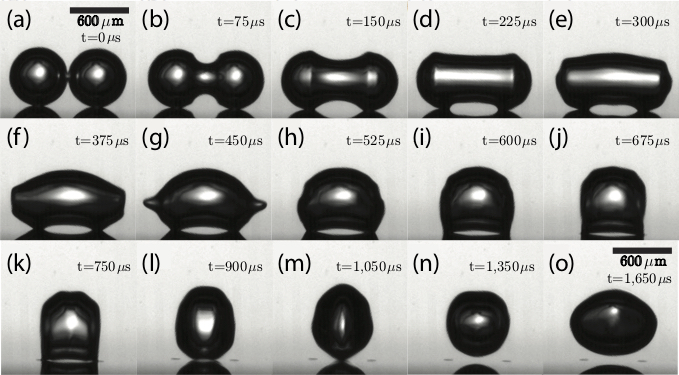}
\caption{\it 
Snapshots of the coalescence of two diffusively grown 
gas bubbles pinned to 50 $\mu$m radius pits. When they touch and coalesce, both bubbles have a
diameter of 600 $\mu$m. 
 Three main stages can be identified: neck formation between the two bubbles, images (a-d); propagation of the capillary waves along the upper and lower half-surface of the coalescing bubble and their final convergence at its side, images (e-l); and final detachment, upwards jumping, rising and oscillation with the Minnaert frequency, images (m-o).
Adopted from ref.\ \cite{moreno2018b}. 
}
\label{moreno}
\end{figure}

An example of a carbon dioxide bubble which is purely growing by diffusion is shown in figure \ref{diff-vs-cat}b. 
As discussed above, according to the Epstein-Plesset theory \cite{epstein1950}, 
for such bubbles the radius should grow in time with a square-root behavior.  As we found 
in ref.\ \cite{enriquez2014}, in the beginning it indeed does, but later on convective effects take over,
as the dissolved carbon dioxide affects the density of water. We also found that in the long term gas 
depletion effects become very relevant and bubbles nucleating later at the same location 
grow much slower than the earlier ones
\cite{moreno2017}. After growth, the bubbles finally detach. This happens because of buoyancy,
either directly once they are so  large that the buoyant forces overwhelm the capillary forces
(the so-called Fritz radius \cite{fritz1935,ogu93}), or when they coalesce and then detach, as shown in figure
\ref{moreno}. In the context of catalysis and electrolysis, the bubble detachment is of course advantageous,
as pointed out above.  

The area of electrolysis and catalysis  
seems to me to be a very important and promising future field for 
 physics of fluids, 
 building a bridge from fluid dynamics to process technology and (colloid) chemistry.

\section{Effective bubble force models, dispersed bubbly flow,  and bubble drag reduction}\label{force}
\begin{figure}[htb]
  \centering
\includegraphics[width=16cm]{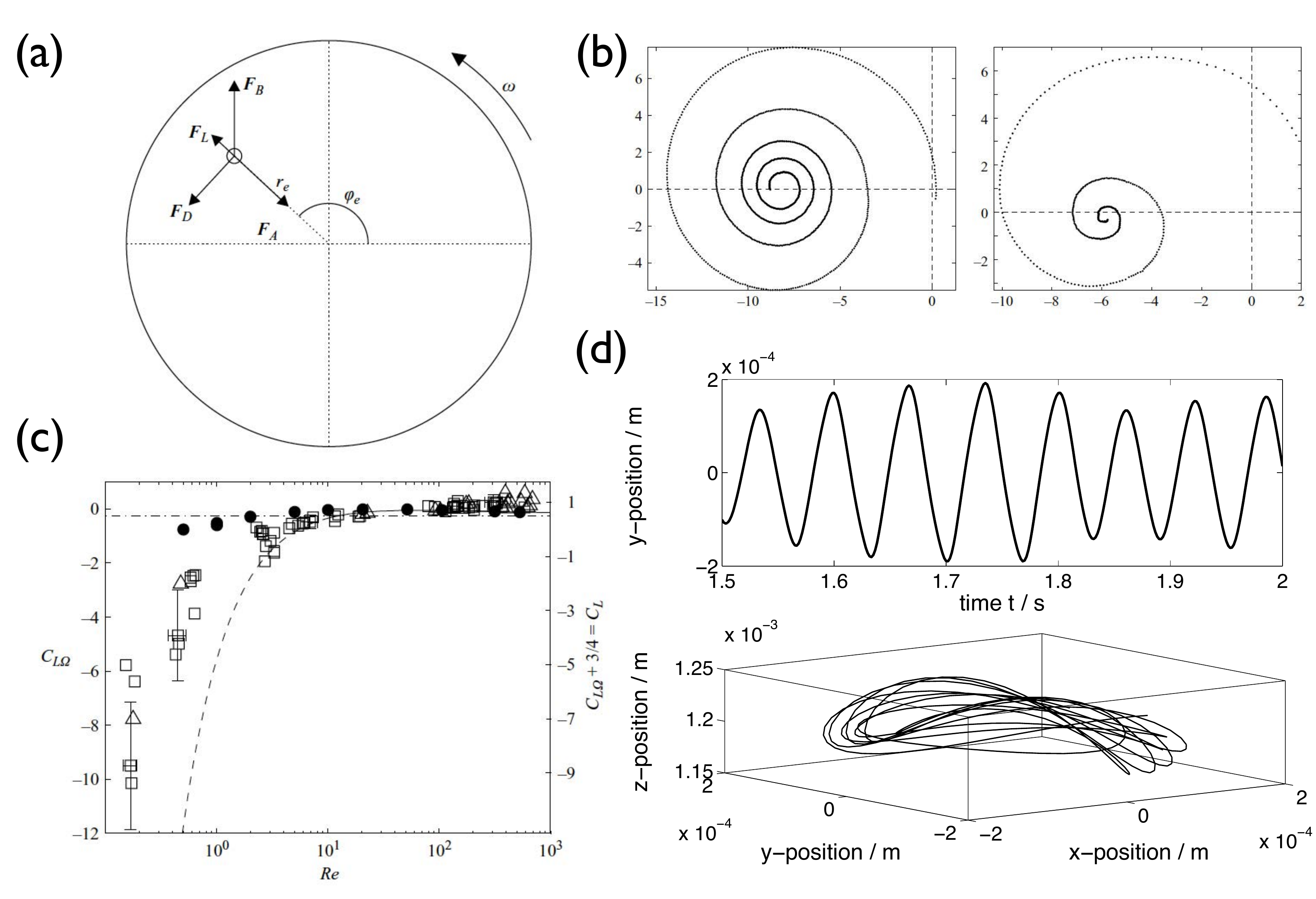}
\caption{\it 
(a) Forces acting on a bubble spiralling in a rotating horizontal cylinder: Drag $F_D$, lift $F_L$, added mass $F_A$, and 
buoyancy $F_B$. The net force typically leads to  a spiralling  trajectory, as shown in 
(b) for bubbles with two different sets of parameters (axes in mm; the center of the cylinder is at (0,0). 
(c) Rotational lift coefficient $C_{L\Omega}$ extracted from such trajectories (open symbols) and compared with
numerical data of ref.\ \cite{mag98}. 
Figure (a) -- (c) are taken from ref.\ \cite{nie07}, to which we refer for more details. 
(d) y-position  and three-dimensional trajectory of a moving SBSL-bubble 
in N-methylformamide \cite{did00b} from a dynamical model based on effective forces (including the history force),
for the same  parameters as in experiment \cite{did00b}, where similar trajectories are observed. 
Taken from ref.\ \cite{toegel2006}.
}
\label{fig-eff-forces}
\end{figure}

I now come back from the predominantly diffusive bubble dynamics to further  inertial bubble phenomena. 
As explained in section \ref{sl}, the acoustic forces (``Bjerknes forces'') are the reason why the  bubble
is pushed towards the pressure antinode where it can oscillate in the strongest way, possibly even leading to 
single bubble sonoluminescence \cite{bre02}. In general, these acoustic forces will compete with  inertial forces 
such as the added mass force, drag, and lift, and of course with buoyancy. In 2001,
we  aimed to understand this competition in more detail, in order to gain more insight into the hydrodynamic
forces on bubbles such as drag and lift 
(which had very nicely been reviewed by Magnaudet and Eames \cite{mag00}) and in order to 
better control the motion of bubbles
in a shear flow with the help of an external acoustic field. We thus built an experimental setup with an acoustic 
 spherical glass resonator 
through which we pumped some liquid  in a 
 controlled way, thus generating   controlled shear flow \cite{ren01}. 
Remarkably, for certain parameters, the entrapped bubbles performed a spiralling motion, out of which we could 
extract various pieces of information on the drag and in particular on the lift. The work -- and follow up work in similarly controlled 
geometries such as a rotating horizontal cylinder \cite{loh03b,nie07} --  also confirmed the usefulness of the concept
of effective bubble forces, even for deformed or oscillating bubbles. In figure \ref{fig-eff-forces}, for a  bubble spiralling in 
the rotating horizontal cylinder \cite{nie07}, we show the forces acting on the bubble (a), the spiralling trajectories for certain parameters (b),
and the extracted rotational lift coefficients $C_L$, which remarkably can take both negative and positive values (c)
\cite{nie07}. 
In a particularly counter-intuitive  example, 
employing the Basset history force as a
non-local-in-time effective force \cite{mag98}, we could explain how viscosity can destabilize
the path dynamics of sonoluminescing bubbles \cite{toegel2006}, a phenomenon discovered by Suslick and coworkers
for sonoluminescing bubbles
in glycol and called moving-SBL \cite{did00b}. The unstable, chaotic  path-trajectory of such a bubble is shown in figure
\ref{fig-eff-forces}d.

\begin{figure}[htb]
  \centering
\includegraphics[width=0.98\textwidth]{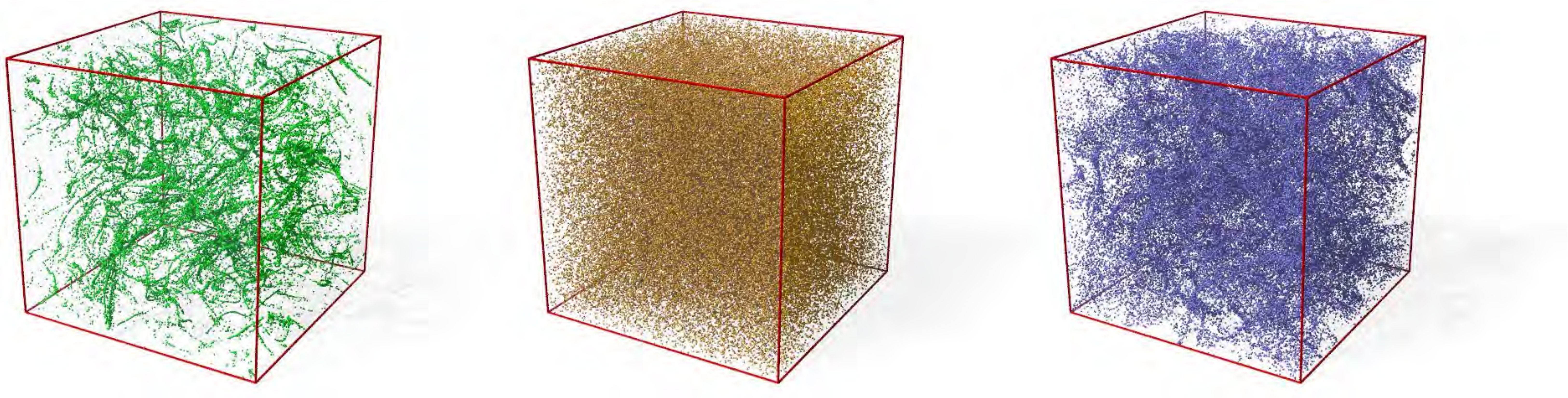}
\caption{\it 
Snapshots of the particle distribution in  turbulent 
flow field for a Stokes number $St=0.6$ for 
(a) bubbles,
(b) tracers, and
(c) heavy particles, all for a Taylor-Reynolds number $Re_\lambda = 75$, obtained with the point-particle approach in the spirit of 
Maxey and Riley \cite{max83}. 
Figure taken
from ref.\
\cite{cal08b}.}
\label{fig-dispersed}
\end{figure}

Encouraged by these successes, we \cite{maz03a,maz03b}
dared to transfer 
Maxey and Riley's seminal idea of using effective forces on small particles to 
describe their (collective) dynamics in dispersed
multiphase flow with {\it point-like particles} 
\cite{max83}  to turbulent dispersed multiphase flow \cite{cro96,bal10,tos09}
with {\it point-like bubbles}, following refs.\ 
\cite{tho84,aut87,wan93b,max94,mag95,sri95,cli96,cli99,spe97,leg98,sri99}.
The motion of the  small and thus non-deformable 
bubbles of radius $a$, volume 
$\calv = 4\pi a^3/3$, velocity $\v(\x ,t)$  and 
embedded in a velocity field ${\bf u} (\x ,t)$ was
 modelled by  (see e.g.\ \cite{tho84,spe97,cli99,mag00})
\begin{eqnarray}
 \rho_g \calv {d\v \over dt} &=& (\rho_g - \rho_f) \calv \g - C_D {\pi a^2 \over 2}
 \rho_f | \v - \u| (\v - \u) \nonumber \\
 &+& \rho_f \calv \left( 
 C_M  {D \u \over D t} - {d\v \over dt}
 \right) 
 + \rho_f \calv {D\u \over Dt } 
- C_L \rho_f \calv (\v - \u ) \times
\bomega 
\label{mr}
\end{eqnarray}
and each bubble at position $\y (t)$ exerted a $\delta$-type force on the liquid flow 
\cite{maz03b},
\begin{equation}
\f_b (\x, t) = \calv \left( 
{D\u \over D t} - \g 
\right)
\delta(\x - \y(t)). 
\label{delta-forcing}
\end{equation}
Here,   $\rho_f$ is the liquid density and 
$\rho_g \ll \rho_f$ is the (low) gas density.
The  terms on the right-hand side of eq.\ (\ref{mr}) represent 
 buoyancy,  drag,   added mass plus fluid acceleration, and lift,
where $C_D$, $C_M$ and  $C_L$ are the corresponding coefficients, which are modelled.  
Though in such complicated turbulent flow situation in particular 
the lift coefficient is not exactly known, as an approximation
-- and in order to explore the effect of the lift force on the overall flow dynamics -- we took the
standard value $C_L =1/2$ \cite{mag00} for small spherical  bubbles in shear flow. 
Following this approach, we studied the effect of the bubbles on the turbulent energy spectra \cite{maz03b},
the 
Lagrangian statistics of bubbles in turbulence \cite{maz04b} finding very pronounced intermittency of the
bubble acceleration, and the clustering of bubbles in the vortices \cite{cal08,cal08b}, see  figure 
\ref{fig-dispersed}, in which we compare the distribution of bubbles with that of particles in dispersed multiphase
flow. 
We also extended the effective bubble force idea to vapor bubbles in turbulent flow 
which can shrink and grow thanks to evaporation and condensation
\cite{ore09,lakkaraju2011,lakkaraju2013}. In this way we could reveal the physics of the considerable enhanced heat
transfer in boiling thermal convection as compared to normal thermal convection.
 Two enhancing effects compete:
Enhancement due to additional mixing by the rising vapor bubbles, and enhancement due 
 to the bubbles
as directed carriers  of (latent) heat from the boiling bottom plate to the cold  top plate. Due to the many modelling
assumptions a quantitative comparison with experimental data \cite{guzman2016} however remains difficult.

\begin{figure}[htb]
  \centering
\includegraphics[width=0.98\textwidth]{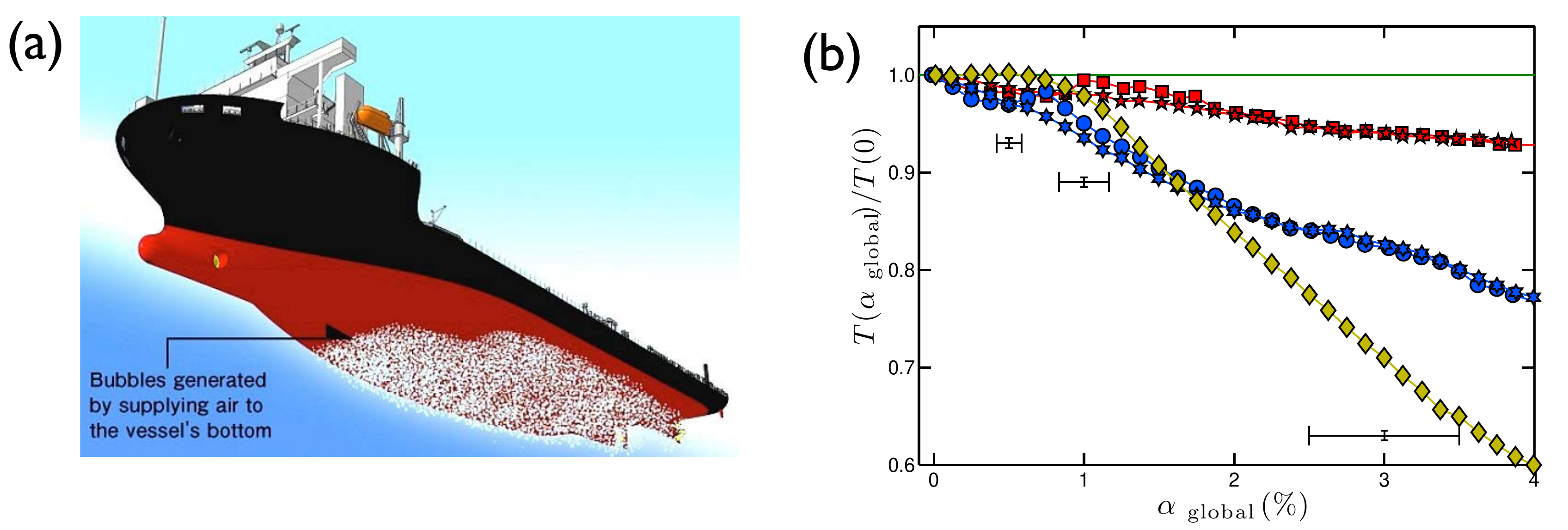}
\caption{\it 
(a) Sketch of the envisioned bubble drag reduction for naval transportation
(Mitsubishi corporation, Japan): Bubbles are injected under the ship hull.
(b) Drag reduction as function of bubble volume concentration in turbulent
bubbly Taylor-Couette flow for three different Reynolds numbers
of the rotating inner cylinder, namely $Re = 0.5 \cdot 10^6$ (red data), $Re = 1.0  \cdot 10^6$ (blue data), 
and $Re = 2.0  \cdot 10^6$ (yellow data). The vertical axis shows the required torque with bubbles divided by the required 
torque in the single flow case. 
Figure 
 taken
from ref.\
\cite{gil13}.
}
\label{bdr-high}
\end{figure}


\begin{figure}[h!]
  \centering
\includegraphics[width=0.92\textwidth]{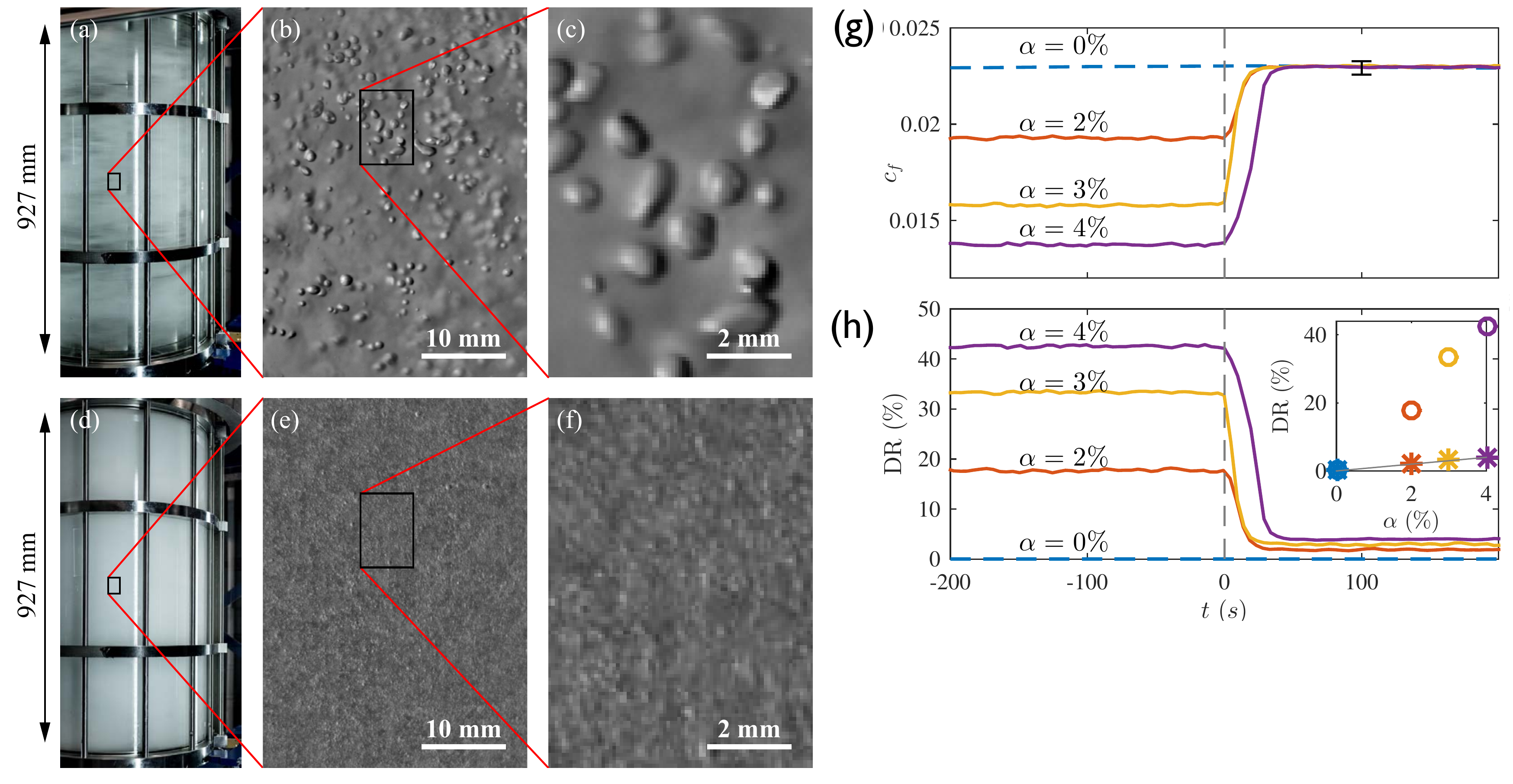}
\caption{\it 
(a-f) Snapshots of bubbly Taylor-Couette
turbulence ($\alpha =1\%$, $Re_i = 2\cdot 10^6$) with increasing magnification (as shown by the scale bars).
In the first row no surfactants (a-c) are present in the turbulent flow, whereas the second row (d-f) shows
the (statistically stationary) situation after addition of 6 ppm Triton X-100. 
In the left photos the Twente Turbulent Taylor-Couette (T$^3$C) apparatus \cite{gils2011}
can be seen.
(g) Friction coefficient $c_f$ as function of time for various bubble volume concentrations $\alpha$. The surfactant is added at time t = 0 s.
(h) Corresponding drag reduction DR (in \%) as function of time. The inset shows how the drag reduction depends on the bubble 
volume concentrations: open circles: without  surfactant; stars: with surfactant. 
Figures taken
from ref.\
\cite{verschoof2016}.}
\label{fig-tc-surf}
\end{figure}


One of the most remarkable features of  dispersed bubbly flow is that a bubble addition of only a few percent
 leads to  remarkable 
drag reduction  of up to 80\% 
\cite{mad84,mad85},  with great potential for naval applications  \cite{kod00drag} (see figure \ref{bdr-high}a), as it can lead to significant reduction of the fuel consumed by ships without adding substances into water. 
As far as I can remember,
I myself first heard  of this phenomenon at an APS DFD meeting in  a talk of Kazu Sugiyama, then a 
postdoc in the group of  Yoichiro  Matsumoto and Shu Takagi in Tokyo/Japan. 
For  excellent recent
reviews on bubble drag reduction  I refer to refs.\ \cite{cec10,murai2014,kumagai2015}. 

Despite all efforts done to understand the fundamental mechanisms behind this effect, a solid understanding of the
drag reduction 
 mechanisms occurring in bubbly flows has  still been 
  missing. To achieve such an understanding, we wanted to study bubble drag reduction 
  under the well-defined 
conditions of turbulent Taylor-Couette (TC) flow, the flow between two coaxial co- or counter-rotating cylinders.  
For recent reviews on low-Reynolds number  and  high-Reynolds number
 TC flow,  we refer to ref.\ \cite{far14} and ref.\ \cite{gro16}, respectively.  
The advantage of this canonical flow of physics of fluids is that it takes place in  a closed system and
that both global measurements (the overall torque required to drive the system, i.e., the drag)  and local measurements
(by PIV or LDV) are possible. In a series of papers on turbulent bubble drag reduction
\cite{ber05,ber07,gil13},  we could indeed measure major bubble drag reduction in this TC system, namely in the strongly
turbulent case up to 40\% with only 4\% volume fraction of the bubbles,
see figure
\ref{bdr-high}b, and in addition characterize bubble concentration and  velocity profiles. The measurements
were consistent with the conjecture that it is the bubble deformability that is responsible for the strong drag reduction,
i.e, we only found substantial drag reduction for bubble Weber numbers We larger than 1 \cite{gil13}. 
These findings were 
in line with the results 
from numerical simulations by 
Gretar Tryggvason's group  \cite{lu05}, employing the front-tracking technique. The modification of the lift force
on the bubbles due to their  deformability (for We $>$ 1) seems to play a central role in the mechanism and
the  bubbles arrange themselves in between the turbulent bulk and the boundary layers, blocking the
momentum transfer and therefore reducing the overall drag. 

How to further validate this hypothesis? If the hypothesis were correct,
one should  change the bubble deformability during the experiment and then the  drag should change
 on the spot. But how to 
achieve this? In 2010 I had visited the group of Yoichiro  Matsumoto and Shu  Takagi in Tokyo and saw
 their impressive
huge 
 bubble column experiment, in which the addition of a few
 drops of surfactants (Triton X) immediately changed
the flow characteristics and bubble size, see ref.\ \cite{takagi2008}; for a recent review of this phenomenon, see 
\cite{takagi2011}. The main reason is that the Triton X surfactant prevents bubble coalescence, and that therefore 
in strongly turbulent flow in which bubble splitting is omnipresent, the average bubble size will dramatically
decrease after the addition of Triton X. 


\begin{figure}[htb]
  \centering
\includegraphics[width=0.88\textwidth]{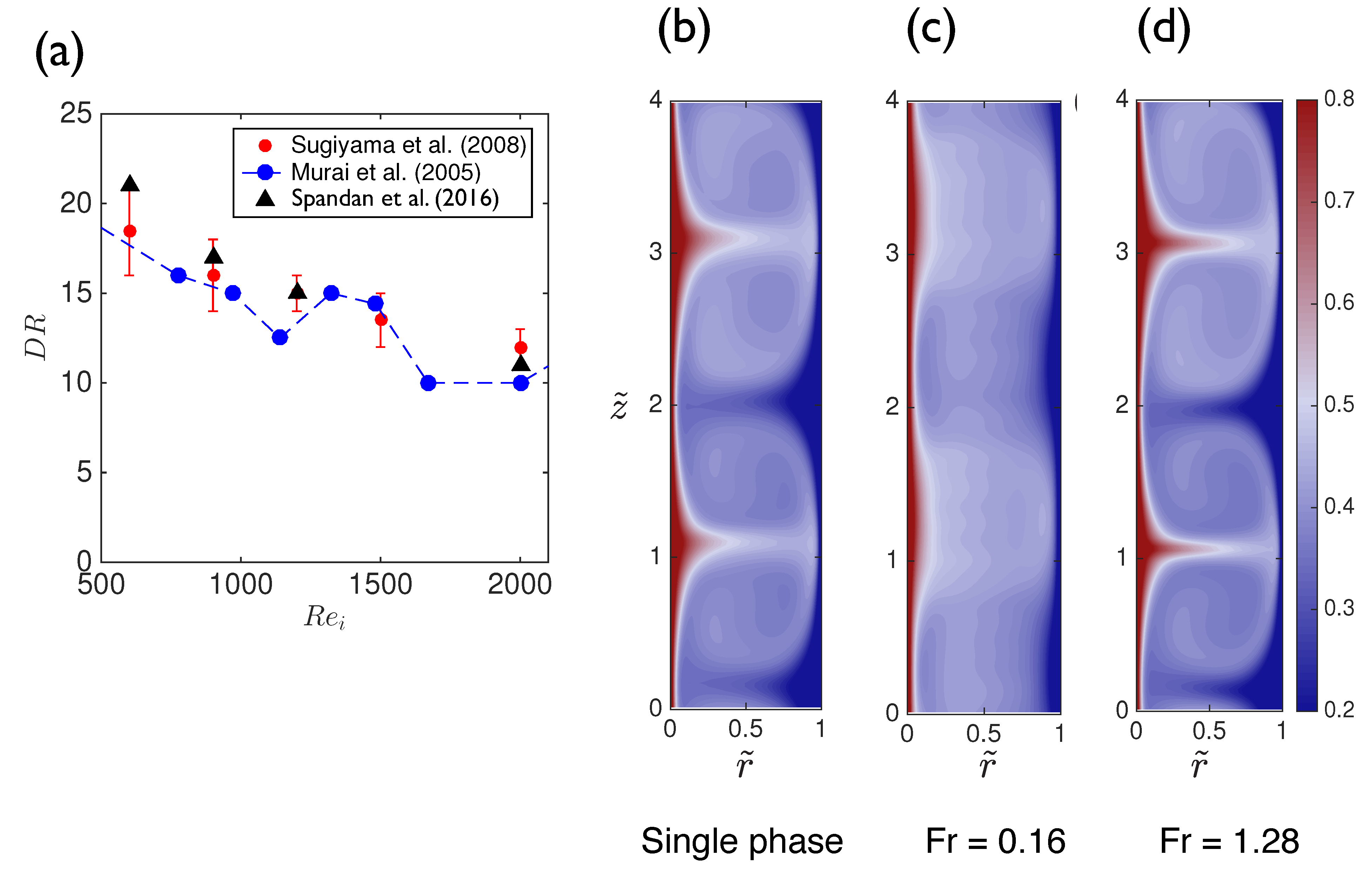}
\caption{\it 
(a) Drag reduction DR as function of the Reynolds number $Re_i$ of the inner cylinder in Taylor-Couette flow 
in the non-turbulent regime: 
The data are from the experiments by 
Murai {\it et al.} \cite{mur05}, the 
numerical simulations of Sugiyama {\it et al.} \cite{sug08b}, 
and the numerical simulations by Spandan {\it et al.} 
\cite{spandan2016a}. One clearly sees that for these low Reynolds numbers the drag reduction is decreasing with increasing
Reynolds numbers. 
(b-d) Corresponding averaged azimuthal velocity contours with point-bubbles included. 
$Re_i =2500$ and single phase flow is compared with two different Froude numbers
$Fr = 0.16$ (strong buoyancy, weakening the Taylor rolls considerable) and $Fr = 1.28$ (weak buoyancy). 
In the latter case, the Taylor rolls are hardly affected by the bubbles which get trapped in them. 
Figures taken
from ref.\
\cite{spandan2016a}.
}
\label{bdr-low}
\end{figure}


We decided to apply the same trick in turbulent bubbly TC flow, and indeed found that also here the 
addition of a few drops of Triton X  (to 111 liter of water in our TC setup \cite{gils2011}) dramatically changed the flow
\cite{verschoof2016}, see figure \ref{fig-tc-surf}, in spite of the unchanged bubble volume fraction: 
Optically  (fig.\ \ref{fig-tc-surf}a-f), because  the flow became very opaque,  just as in ref.\  \cite{takagi2008},
but in particular also mechanically (fig.\ \ref{fig-tc-surf}g-h), as  the effect of bubble drag reduction nearly vanished.
The combination of these local  and global  findings
 strongly supports the hypothesis that it is the bubble deformability which is relevant for strong
turbulent bubble drag reduction. 

How to numerically model turbulent bubble drag reduction? Given that the mechanism is bubble deformability, 
which requires to resolve the gas-liquid interfaces of the bubbles and given the large Reynolds numbers, this is
extremely challenging. With the front-tracking technique, Tryggvason and coworkers  could simulate
a few large bubbles in a ``minimum turbulent  channel'', indeed qualitatively finding bubble drag reduction 
due to deformability \cite{lu05}. 
Also his later work with more bubbles confirmed this interpretation  \cite{lu2008,dabiri2013}. 

For the numerical simulations of bubble drag reduction in TC flow, we started off moderately, namely for relatively small Reynolds numbers (based on the inner cylinder rotation, the outer cylinder was
 at rest) $Re_i < 5000 $, for which the flow is not yet turbulent. 
Note that these Reynolds numbers are more than two orders of magnitude smaller than what 
I discussed in figures \ref{bdr-high} and \ref{fig-tc-surf}. 
Interestingly enough, also  in that low Re-regime Murai et al.\ \cite{mur05} had experimentally 
 found  bubble drag reduction, namely  by microbubble
injection, see figure \ref{bdr-low}a. But due to the lack of bubble deformability (We $<$ 1 for the injected 
microbubbles), the physical 
mechanism in that regime must be very different. The good news is that this regime is accessible to 
numerical simulations, and even with effective force models, as the microbubbles can be 
considered as  point-like, with $We < 1$ and a diameter  $d_b < 10 \eta$, where $\eta$ is the (global) Kolmogorov length
of the flow.


\begin{figure}[htb]
  \centering
\includegraphics[width=0.98\textwidth]{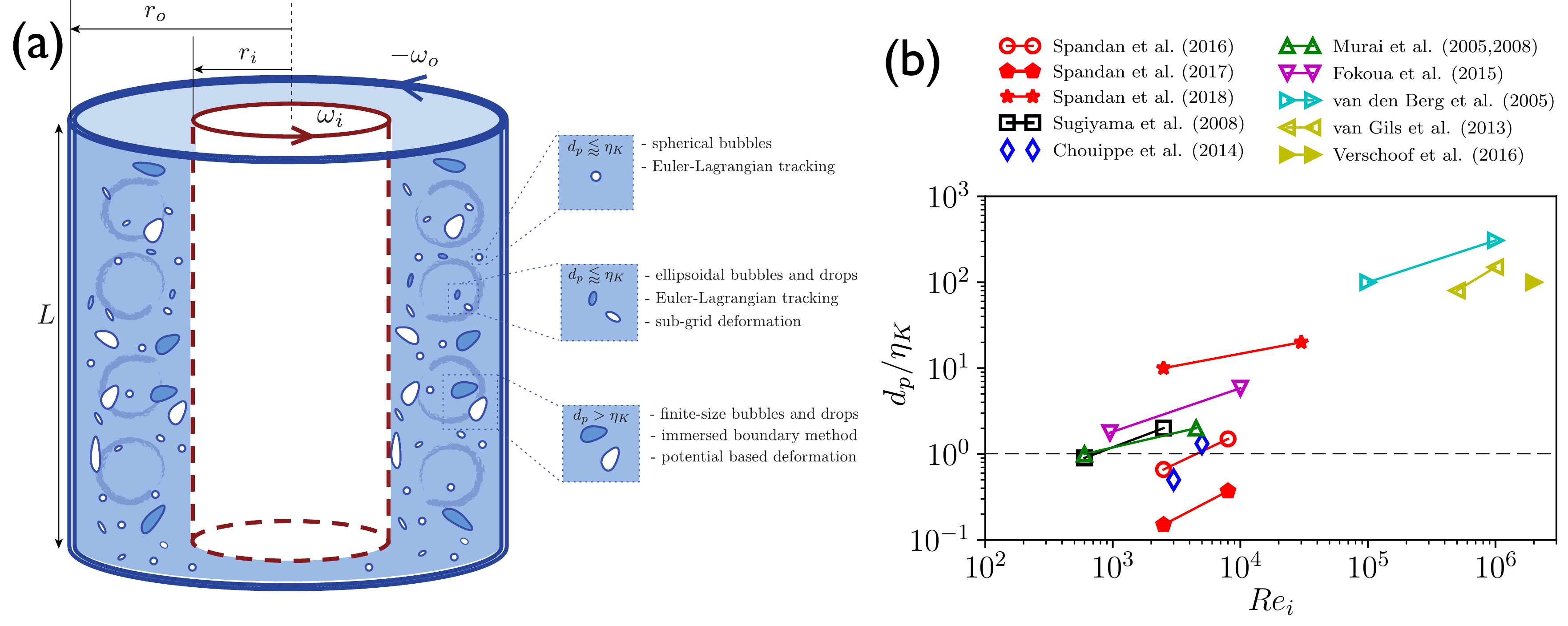}
\caption{\it 
(a) Visualization of the different ways to numerically treat bubbles in TC flow, depending on their
sizes. 
 Figure adopted from the PhD thesis of Vamsi Spandan, Physics of Fluids, University of Twente, 2017. 
(b)
Parameter space of bubble drag reduction in TC flow: Various bubble sizes $d_b$ (given 
 relative to the Kolmogorov scale $\delta_\nu$) were realized in experiments and simulations for Reynolds numbers $Re_i$ from a few 
 hundreds up to more than $10^6$. The left block of data are experiments and numerical simulations for 
 microbubble drag reduction from refs.\ \cite{mur05,sug08b,chouippe2014,fokoua2015}, and the right block of data are our large Re 
 experiments 
 with deformable bubbles \cite{ber05,ber07,gil13}. 
}
\label{vamsi1}
\end{figure}

\begin{figure}[htb]
  \centering
\includegraphics[width=0.8\textwidth]{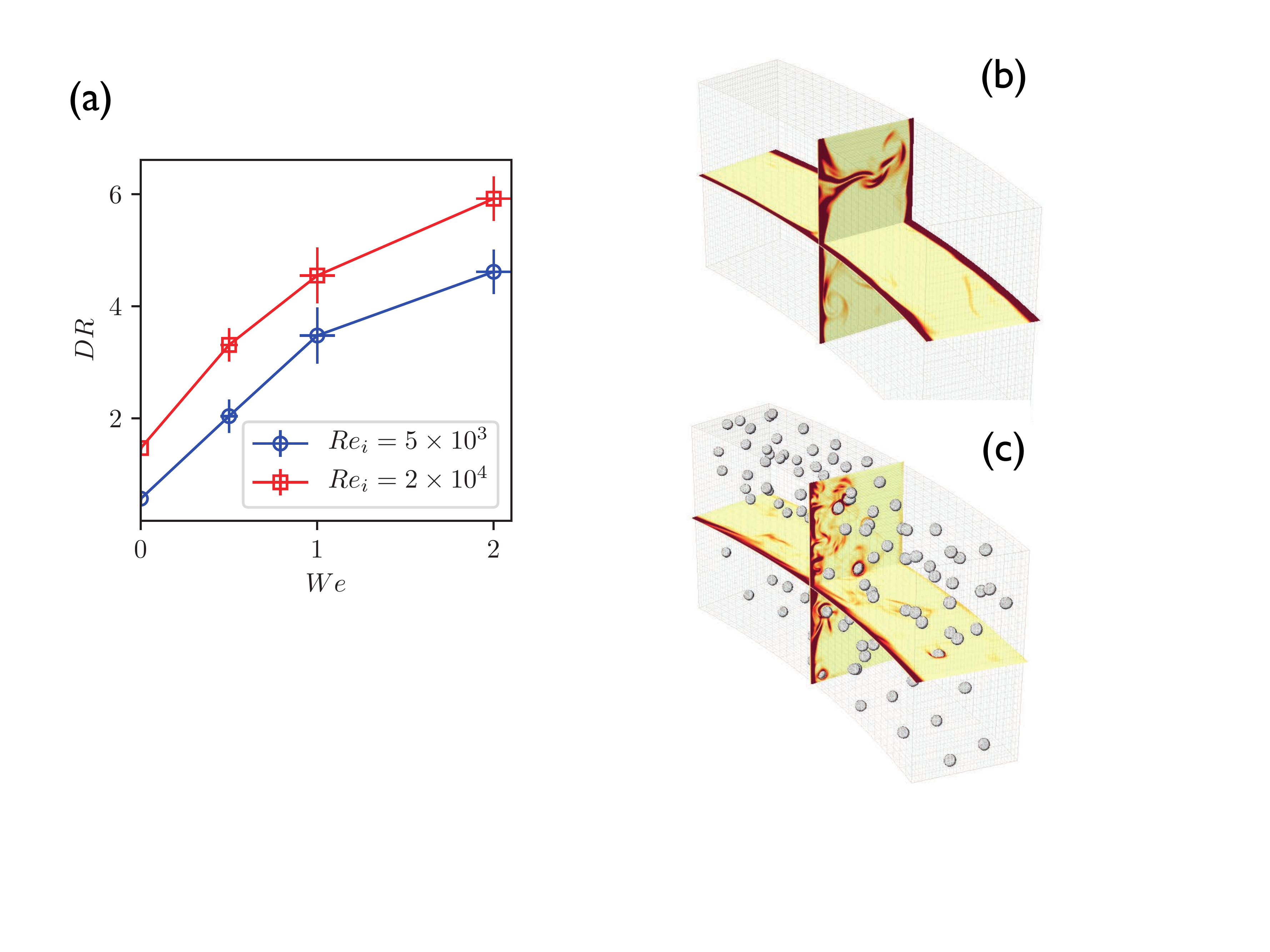}
\caption{\it 
(a) Drag reduction in turbulent TC flow as function of Weber number for two different Reynolds numbers:
Increasing Weber number implies increasing bubble deformability which is crucial for the drag reduction.
Taken from the numerical simulations of ref.\  \cite{spandan2018}.
 (b,c) Numerical simulations showing plume detachment from the inner TC cylinder for single phase flow (b) and bubbly flow (c), where the momentum transfer is
 considerably depressed. In the latter case deformable bubbles  considerably lower the momentum transfer. 
 The calculation was done with the immersed boundary method.  
}
\label{vamsi2}
\end{figure}


Our numerical  simulations of bubbly TC flow at these  low $Re_i < 5000$ and with point-like bubbles 
\cite{sug08b,spandan2016a}  indeed resulted in  bubble 
 drag reduction in quantitative agreement with Murai's data \cite{mur05},
see figure \ref{bdr-low}a. In contrast to the bubble drag reduction in the large
 Reynolds number regime, which increases with increasing 
  Reynolds number (figure \ref{bdr-high}b), in the low Reynolds
 number regime the effect of bubble drag reduction gets smaller with increasing Reynolds number
  (figure \ref{bdr-low}a).

As the relevant mechanism for bubble drag reduction in TC flow in the low Reynolds number regime,
we could identify  the considerable weakening  of the Taylor rolls through the rising bubbles, see 
figures \ref{bdr-low}b-d. The crucial parameter in this low Reynolds number regime is therefore the
Froude number (and not the Weber number as in the large Reynolds number regime), 
here defined as square root of the ratio between
centrifugal force and buoyancy, 
$Fr = \omega_i \sqrt{r_i/g}$, where $r_i$ is the bubble radius, $\omega_i$ the angular
velocity of the inner cylinder and $g$ gravity. For small $Fr<1$ 
 (figure \ref{bdr-low}c), buoyancy wins, and the bubbles rise through the 
Taylor rolls, thereby weakening them. This reduces the angular momentum transfer from inner to outer cylinder
and thus the drag. For large $Fr> 1$ 
 (figure \ref{bdr-low}d) the bubbles simply get trapped in the Taylor rolls, without affecting them much, and the
 overall drag remains nearly unchanged \cite{spandan2017b}. 
 For increasing Reynolds numbers the 
 Taylor rolls  more and more lose their flow dominance which explains why in this low Reynolds number
 regime the bubble drag reduction decreases with increasing Reynolds number (figure \ref{bdr-low}a).
 
 The challenge for the next years will be to close the gap between the microbubble drag reduction experiments
 at lower Re of Murai et al. \cite{mur05} and our large Re experiments with deformable bubbles \cite{gil13}, see
 figure \ref{vamsi1}.
Our first step on this way was to include bubble deformability in point-particles by picking up an idea of 
refs.\ \cite{mm1998,biferale2014} and couple the deformation dynamics of  ellipsoidal bubbles
to the flow. Indeed, we found that bubble deformation helped the overall drag reduction 
\cite{spandan2017b}. In a second step \cite{spandan2018}, we coupled the well-resolved 
 bubbles with an immersed boundary method 
to the Navier-Stokes equation, again finding that 
 an increase in the bubble deformability (i.e., its Weber number) indeed 
 implies larger  drag reduction, see figure \ref{vamsi2}a. 
  Profiles of the 
   local angular velocity flux show that, in the presence of bubbles, turbulence is enhanced near the inner cylinder while attenuated in the bulk and near the outer cylinder. We connect the increase in drag reduction to the decrease in dissipation in the wake of highly deformed bubbles near the inner cylinder \cite{spandan2018},
  as visualized in figure \ref{vamsi2}b,c. As in the closed TC system the total dissipation is proportional to the 
  total torque \cite{gro16}, the decrease in dissipation implies a decrease in the overall drag. 
  
  Bringing all 
  these fundamental 
   insights into bubble drag reduction towards 
    applications in naval industry however still is a very long way 
  to go and other concepts like air cavities may be more efficient \cite{verschoof2018b}.
We are presently collaborating with the Dutch Maritime Research Institute (MARIN)  in Wageningen on this subject.

\section{Conclusions}\label{conclu}
I hope to have demonstrated -- along my own bubble trajectory -- how fascinating and broad
the fluid dynamics  of a bubble is, from the nanometer scale to geophysical scales,
with both outstanding fundamental questions and relevant applications. Clearly, 
the close interaction between experiment, theory, and numerics
is crucial in physics of fluids and in particular in physics of bubbles, and makes this subject so particularly nice.
Only a combination of all three methods gives real insight into the
phenomena. 

I also hope to have demonstrated with the chosen examples how
crucial and essential for innovation of any kind  {\it fundamental} research is. 
In what direction creative fundamental research will go is often
unpredictable. If one ends up exactly where one had expected, the
chance is large that the research was standard. What remains crucial is to be open, watch, and listen. 

I want to close with some more general comments on fluid dynamics: From my point of view we presently 
live in the golden age of fluid dynamics: The reasons are
 that (i) Moore's law is kept on still being followed
for the computational power, now making simulations possible of which even ten years ago we did not dare to dream of, and
(ii) a similar revolution (for the same reason) in digital 
high-speed imaging, now being able to routinely resolve  the millisecond time scale and even smaller time scales \cite{versluis2012}, revealing new physics on these scales which up to now was inaccessible
and producing a huge amount of data on the flow. 
Also other advanced equipment like confocal microscopy, digital holographic microscopy and atomic force microscopy get more and more used
in fluid dynamics. 
With all  of these advances 
together, the gap between what can be measured and what can be ab-initio simulated is more quickly closing 
than we had anticipated at the end of the last century. 

Also other gaps are closing: Fluid dynamics and in particular that of bubble fluid dynamics 
 is bridging out to various neighboring disciplines such as chemistry and in particular colloidal science, catalysis, electrolysis, medicine, biology, computational science, and many others. Here the techniques, approaches and traditions from fluid dynamics can offer a lot to help to solve outstanding problems --
 vice versa, these fields can offer wonderful questions to fluid dynamics. 
Academic (bubble) fluid dynamics is also bridging out not only to traditional applications on large scales such as in chemical engineering, in the food industry, or in geophysics, but also to various new high-tech applications, be it in inkjet printing, immersion and XUV lithography, chemical diagnostics, and lab-on-a-chip microfluidics.

\section*{Acknowledgement} 
Science is a collective activity. The scientific insight in this paper is
 gained during my bubble journey over more than
two decades 
 together with my colleagues, postdocs, PhD students, and students and I would like to thank them for all their work and contributions, and for the 
stimulation and intellectual pleasure we have enjoyed when doing physics. 
But first of all, I would like to express my sincere gratitude to my scientific teacher, Professor Siegfried Grossmann, for all what I learnt from him,
which goes way beyond physics. 
Then, in 
particular, on this subject of bubbles, I would like to thank Michael Brenner,
Sascha Hilgenfeldt and Michel Versluis, with whom my bubbly journey started, and 
the  ``part-time professors" of our Twente Physics of Fluids group, namely my 
colleagues and friends
Andrea Prosperetti (who also inspired us on the beauty of bubbles in art \cite{pro04}!), 
Roberto Verzicco,  Chao Sun, and Xuehua Zhang,
whose intellectual contribution to our group is way beyond ``part-time'', from whom I learned  tremendously,
and with whom it has been  a great pleasure and privilege to work. 

I also gratefully acknowledge generous financial support over the years, mainly from NWO via various programs and from the
 European Union via ERC Advanced Grants.

\bibliography{}  

\end{document}